\documentclass[review, sort&compress, 3p]{elsarticle}

\usepackage{lineno,hyperref}

\graphicspath{{figures/}}
\usepackage{amsmath,amsfonts,amssymb}
\usepackage{bm}

\usepackage[skip=0pt]{caption}
\usepackage[inline]{enumitem}
\usepackage{subcaption}
\usepackage{multicol}
\usepackage{xcolor}
\usepackage[normalem]{ulem}
\usepackage[nameinlink,capitalize]{cleveref}
\begin{document}
	
	\begin{frontmatter}
		
		\title{A metric for assessing and optimizing data-driven prognostic algorithms for predictive maintenance}
		
		\author[mymainaddress,mysecondaryaddress,mytertiaryaddress]{Antonios Kamariotis \corref{mycorrespondingauthor}}
		\cortext[mycorrespondingauthor]{Corresponding author}
		\ead{antonis.kamariotis@tum.de}
		\author[mysecondaryaddress]{Konstantinos Tatsis}
		\ead{tatsis@ibk.baug.ethz.ch}
        \author[mysecondaryaddress,mytertiaryaddress]{Eleni Chatzi}
		\ead{chatzi@ibk.baug.ethz.ch}
   	\author[myquaternaryaddress]{Kai 
        Goebel}
 	\ead{kgoebel@parc.com} 
		\author[mymainaddress]{Daniel Straub}
		\ead{straub@tum.de}

		\address[mymainaddress]{Engineering Risk Analysis Group, Technical University of Munich, Theresienstrasse 90, 80333 Munich, Germany}
		\address[mysecondaryaddress]{Institute of Structural Engineering, ETH Zurich, Stefano-Franscini-Platz 5, 8093 Zurich, Switzerland}
		\address[mytertiaryaddress]{Institute for Advanced Study, Technical University of Munich, Lichtenbergstrasse 2a, 85748 Garching, Germany}
 		\address[myquaternaryaddress]{Palo Alto Research Center (PARC), Palo Alto CA 94304, USA}
		
        \begin{abstract}
        Prognostic Health Management aims to predict the Remaining Useful Life (RUL) of degrading components/systems utilizing monitoring data. These RUL predictions form the basis for optimizing maintenance planning in a Predictive Maintenance (PdM) paradigm. We here propose a metric for assessing data-driven prognostic algorithms based on their impact on downstream PdM decisions. The metric is defined in association with a decision setting and a corresponding PdM policy. We consider two typical PdM decision settings, namely component ordering and/or replacement planning, for which we investigate and improve PdM policies that are commonly utilized in the literature. All policies are evaluated via the data-based estimation of the long-run expected maintenance cost per unit time, using monitored run-to-failure experiments. The policy evaluation enables the estimation of the proposed metric. We employ the metric as an objective function for optimizing heuristic PdM policies and algorithms' hyperparameters. The effect of different PdM policies on the metric is initially investigated through a theoretical numerical example. Subsequently, we employ four data-driven prognostic algorithms on a simulated turbofan engine degradation problem, and investigate the joint effect of prognostic algorithm and PdM policy on the metric, resulting in a decision-oriented performance assessment of these algorithms.
        \end{abstract}
		
		\begin{keyword}
			PHM, predictive maintenance, data-driven prognostics, RUL, performance metrics, CMAPSS
			%\texttt{elsarticle.cls}\sep \LaTeX\sep Elsevier \sep template
			%\MSC[2010] 00-01\sep  99-00
		\end{keyword}
		
	\end{frontmatter}
	
	% \linenumbers

  %%%%%%%%%%%%%%%%%%%%%%%%%%%%%%%%%%%%%%%%%%%%%%%%%%%%%%%%%%%%%%%%%%%%%%%%%%%%%%%%%%
	\section{Introduction}
	\label{sec:introduction}
	
    Prognostics is primarily concerned with the prediction of the time instance when a system or component loses its functionality \cite{Kim_2017}. In the context of an on-line monitoring strategy, this typically implies the use of a prognostic model, which furnishes estimates of the Remaining Useful Life (RUL) on the basis of available measurements of the system's response \cite{Galar_2021}. The subsequent task of optimal maintenance planning informed by the prognostic model output is known as health management. Prognostic Health Management (PHM) \cite{Kim_2017, Galar_2021, Zio_2022} is the umbrella term used to define this procedure. Multiple sources of uncertainty enter the prognostics process, which motivates the adoption of a stochastic approach to the estimation of the RUL \cite{Fink_2020}. Consequently, the associated maintenance planning can be defined as a sequential decision problem under uncertainty \cite{Kochenderfer_2015,PAPAKONSTANTINOU2014202,MEMARZADEH2016137,Luque_2019,Arcieri_2022}.
	
    One can distinguish between model-based and data-driven prognostic methods \cite{Kim_2017}. A recent review paper \cite{Lei_2018} classified prognostic approaches into four distinct categories: physics model-based approaches, statistical model-based approaches, artificial intelligence (AI) approaches and hybrid approaches. An overview of recent literature reveals the increasing popularity of data-driven AI approaches \cite{Li_2018, Wu_2018, Li_2019}, owing amongst other factors, to their applicability in case where the degradation pattern cannot be easily represented a-priori via physics-based or statistical models \cite{Nectoux_2012,Lei_2018}. On the downside, it is often not straightforward to capture the uncertainty in machine learning (ML) predictions \cite{Fink_2020, Nguyen_2022}. The task of quantifying such an uncertainty, that is inherent in RUL predictions, is essential for subsequent maintenance planning tasks.
	
    The PHM community has established various experimental and numerical prognostic datasets, which typically contain multivariate time series data obtained from continuous monitoring of run-to-failure experiments on deteriorating components/systems, such as rolling bearings \cite{Nectoux_2012}, batteries \cite{Saha_2007}, turbofan engines \cite{CMAPSS} and industrial machines \cite{purohit2019mimii}. Several of these datasets have been made publicly available by the NASA Prognostics Center of Excellence \cite{NASA}. The availability of such datasets has paved the way for the development and training of a multitude of data-driven prognostic algorithms. These are reviewed in \cite{SI20111,Javed_2017,Lei_2018}. Most of the available literature focuses on the RUL prediction task \cite{Li_2018, Wu_2018,Li_2019,ARIASCHAO2022107961}, and does not consider the subsequent health management task.
	
    For the task of health management on the basis of RUL predictions, the predictive maintenance (PdM) paradigm stands out \cite{Goebel_2021,Compare_2019,Fink_2020_b}. PdM tasks usually relate to planning intermittent inspections and maintenance \cite{Do_2015, KIM2022108391}, and planning maintenance actions informed via continuous monitoring \cite{DePater_2022,Nguyen_2019, YAN2022109053}. PdM can be classified as either model-based PdM or data-driven PdM \cite{Nguyen_2019, Fink_2020_b}. The former is based on the assumption that a physics-based model, e.g., the Paris-Erdogan law for fatigue crack growth \cite{Paris_1963}, or a statistical process model, e.g., a Gamma process \cite{VANNOORTWIJK_2009} or a Wiener process \cite{ZHANG2018775}, is available for describing the deterioration process. The performance of model-based PdM depends on the adopted model. Most PdM studies to date either employ model-based PdM \cite{Do_2015, Huynh_2018,Mancuso_2021, DEPATER2021107761, YAN2022109053, KAMARIOTIS2022108465}, or simplistically consider hypothetical models of the prognostics information and only focus on the maintenance decision optimization \cite{Fauriat_2019, Benaggoune_2020}. End-to-end data-driven PdM frameworks (from data-driven prognostics to data-driven PdM planning) have recently been introduced and applied on prognostic datasets \cite{Nguyen_2019, DePater_2022, Chen_2022, MITICI2023109199, ZHUANG2023109181}. The data-driven PdM framework relies on availability of a sufficient amount of monitoring data from run-to-failure experiments. These are required both for the training of data-driven prognostic algorithms as well as for the data-driven evaluation of PdM policies. 
	
    Given the diversity of prognostic models, the definition of metrics for assessing and comparing the performance of prognostic algorithms can be of defining importance in decision making \cite{Saxena_2008, Saxena_2010, Nectoux_2012, Javed_2015, Hu_2016, Atamuradov_2017, DePater_2022b}. Recent papers review such metrics \cite{Lei_2018, LEWIS2022108473}. Most of these metrics only implicitly account for the subsequent health management task in their design, e.g., how early the algorithm allows for prediction \cite{Saxena_2010, Nectoux_2012}. Our conjecture is that the choice of a performance metric should be guided by the type of PdM decisions that are to be triggered by the algorithms' outcome. 
	
    The current paper proposes a metric for assessing the efficacy of data-driven prognostic models based on their impact on downstream PdM decisions. We clarify the role that PdM policies play in the definition of this metric. The metric can be applied within any given decision setting. In this paper, two PdM decision settings are considered, namely i) component replacement planning and ii) component ordering-replacement planning, which are fairly common for industrial components. We thoroughly investigate some PdM policies of different complexity that are utilized in the literature. These dynamically receive as input the RUL prediction from prognostics and opt for the actions that should lead to an optimal balance between the predicted risk of failure/risk of late order for a new component, and the benefit of extending the life-cycle of a component/not keeping a spare component in the inventory, respectively. Alternatives and improvements to these PdM policies are proposed. We evaluate these in terms of the estimation of the long-run expected maintenance cost per unit time \cite{Do_2015, Fauriat_2020}, upon applying the policies on a run-to-failure dataset. The proposed metric is evaluated on the basis of this estimation.
    
    The paper is organized as follows. \cref{sec:PdM} introduces the proposed decision-oriented metric and discusses its data-driven estimation via samples of run-to-failure experiments. Subsequently, the metric is described within the context of two typical PdM decision settings, for which specific PdM policies are presented and discussed. \cref{sec:Generic_example} introduces a virtual RUL simulator, which serves as an initial test-bed for investigating different aspects of optimality/sub-optimality of the presented PdM policies and their effect on the metric. \cref{sec:Case_study} contains numerical investigations on an actual case study related to degrading turbofan engines, by use of the well-known CMAPSS prognostic dataset \cite{CMAPSS}. Four different data-driven prognostic algorithms, three classifiers and one regression model, are implemented and compared on the basis of the proposed decision-oriented metric. The interplay between RUL prediction algorithm and PdM policy on the metric is also investigated. Finally, \cref{sec:Conclusions} discusses and concludes this work. 
 %%%%%%%%%%%%%%%%%%%%%%%%%%%%%%%%%%%%%%%%%%%%%%%%%%%%%%%%%%%%%%%%%%%%%%%%
	\section{Predictive maintenance decision policies on the basis of RUL predictions}
	\label{sec:PdM}
	
    To evaluate the quality of different prognostic algorithms that deliver RUL predictions, we compare their performance on subsequent PdM decision making. To evaluate the RUL-based decisions, policies are introduced \cite{Jensen_2007}. A policy is a rule that determines the action to take at time $t$, based on the available information up to that time, i.e., past monitoring data and performed actions. For example, a policy answers the following question: ``Preventively replace the component?" \{yes, no\}. In this work, we consider only decision settings in which the policy is the same at all times, and we refer to this stationary policy \cite{Kochenderfer_2015} as the \textit{PdM policy}.

    \subsection{Decision-oriented metric for prognostics performance evaluation}

    Towards the goal of providing a formal decision-oriented framework for assessing and optimizing the performance of prognostic algorithms, this section proposes a metric that quantifies the optimality of the resulting maintenance decisions triggered by the algorithm's RUL predictions within any given decision setting. The proposed framework is summarized in \cref{fig:synoptic_scheme}.
    
    \subsubsection{Data-based evaluation of a generic PdM policy}
    \label{subsubsec:Data_based_evaluation_generic}
     The long-run expected maintenance cost per unit time (over an infinite time horizon) is typically the quantity of interest when evaluating a PdM policy. According to renewal theory \cite{Tijms, Do_2015, Fauriat_2020}, specifically the renewal-reward theorem, the long-run expected maintenance cost per unit time corresponds to the ratio:
	\begin{equation}
	R=\dfrac{\text{E}[C_{\mathrm{m}}]}{\text{E}[T_{\mathrm{lc}}]},
	\label{eq:ratio}
	\end{equation}
	where $\text{E}[C_{\mathrm{m}}]$ is the expected maintenance cost induced within one life-cycle of the component when following a certain PdM policy, and $\text{E}[T_{\mathrm{lc}}]$ is the expected length of one life-cycle. This result is valid without discounting \cite{VANNOORTWIJK_2009}; for renewal theory with discounting, see, e.g., \cite{PANDEY201727}. It allows evaluating the above expectations for a PdM policy by applying it on $n$ independent single life-cycles. In the vast majority of the literature, this evaluation is done with the aid of a model, with which a large number of life-cycle realizations is simulated \cite{Do_2015}. In cases where data from multiple run-to-failure experiments are available, the expectations in \cref{eq:ratio} can be evaluated based on the data. More specifically, a PdM policy can be applied on $n$ independent components of the same type (e.g., $n$ engines of the same type in the CMAPSS dataset \cite{CMAPSS}) and the cost of maintenance $C_\mathrm{m}$ and the lifetime $T_\mathrm{lc}$ (also known as a renewal cycle in the context of renewal theory \cite{Tijms}) of each component can be evaluated. The expectations in the numerator and denominator of \cref{eq:ratio} can then be approximated as:
	\begin{equation}
		\dfrac{\text{E}[C_{\mathrm{m}}]}{\text{E}[T_{\mathrm{lc}}]} \approx \hat{R} =\dfrac{\frac{1}{n}\sum_{i=1}^n C_{\mathrm{m}}^{(i)}}{\frac{1}{n}\sum_{i=1}^n T_{\mathrm{lc}}^{(i)}},
		\label{eq:ratio_MCS}
	\end{equation}
	where $\hat{R}$ denotes the data-based estimator of the quantity in \cref{eq:ratio}, $C_{\mathrm{m}}^{(i)}$ and $T_{\mathrm{lc}}^{(i)}$ are the cost of maintenance and the lifetime of the $i$-th component, respectively. The lifetime $T_\mathrm{lc}^{(i)}$ is the time to failure or replacement of the component.
	
	With finite $n$, the estimate of $R$ provided by \cref{eq:ratio_MCS} is subject to uncertainty. A first-order approximation of the variance of the estimator in \cref{eq:ratio_MCS} is \cite{Kempen_2000}:
	
	\begin{equation}
        \text{Var}[\hat{R}] \approx \frac{1}{n}\left[\frac{\text{Var}[C_{\mathrm{m}}]}{\text{E}[T_{\mathrm{lc}}]^2} + \frac{\text{E}[C_{\mathrm{m}}]^2 \cdot \text{Var}[T_{\mathrm{lc}}]}{\text{E}[T_{\mathrm{lc}}]^4} - 2 \cdot \frac{\text{E}[C_{\mathrm{m}}] \cdot \text{Cov}[C_{\mathrm{m}}, T_{\mathrm{lc}}]}{\text{E}[T_{\mathrm{lc}}]^3}\right].
		\label{eq:variance_ratio_MCS}
	\end{equation}
	
	It is noted that occasionally in literature the expectation of the ratio $\text{E}\left[\dfrac{C_{\mathrm{m}}}{T_{\mathrm{lc}}}\right]$ is evaluated, instead of the ratio of the expectations of \cref{eq:ratio,eq:ratio_MCS}. This is only an approximation, which can be poor if the variance of the denominator is large.
 
    \subsubsection{Data-based evaluation of the perfect PdM policy}
    
    As a reference, we consider the hypothetical scenario of perfect prognostics, in which the time to failure is known exactly. Perfect prognostics would lead to perfect PdM decisions for each component. Based on the $n$ run-to-failure experiments, the long-run expected maintenance cost per unit time of the perfect PdM policy can be estimated via \cref{eq:ratio_perfect_PdM}:
	\begin{equation}
		R_\mathrm{perfect}=\frac{\text{E}[C_\mathrm{m,perfect}]}{\text{E}[T_\mathrm{lc, perfect}]} \approx \hat{R}_\mathrm{perfect} = \dfrac{\frac{1}{n}\sum_{i=1}^n C_{\mathrm{m,perfect}}^{(i)}}{\frac{1}{n}\sum_{i=1}^n T_\mathrm{lc,perfect}^{(i)}},
		\label{eq:ratio_perfect_PdM}
	\end{equation}
    where $C_{\mathrm{m,perfect}}^{(i)}$ and $T_{\mathrm{lc,perfect}}^{(i)}$ are the optimal cost of maintenance and length of the first life-cycle of the $i$-th component, respectively.  
    
    \subsubsection{Proposed metric}
    To specify the decision-oriented metric for assessing prognostic algorithms, we evaluate the long-run expected maintenance cost per unit time that is achieved with a specific prognostic algorithm in combination with a PdM policy, which is then compared with the respective quantity obtained from perfect prognostics. We define a scalar metric $M$ as the relative difference between the two, based on which the performance of a prognostic algorithm can be assessed with respect to the PdM decisions that are triggered by its outcome:
    \begin{equation}
		M = \dfrac{\dfrac{\text{E}[C_\mathrm{m}]}{\text{E}[T_\mathrm{lc}]} - \dfrac{\text{E}[C_\mathrm{m,perfect}]}{\text{E}[T_\mathrm{lc, perfect}]}}{\dfrac{\text{E}[C_\mathrm{m,perfect}]}{\text{E}[T_\mathrm{lc, perfect}]}}.
		\label{eq:metric}
	\end{equation}
  Based on the $n$ run-to-failure experiments, the metric $M$ can be estimated as:
	\begin{equation}
		\hat{M} = \dfrac{\hat{R}-\hat{R}_\mathrm{perfect}}{\hat{R}_\mathrm{perfect}}.
		\label{eq:metric_hat}
	\end{equation}
	$M=0$ is the optimal result; the larger the value of the metric $M$, the worse the performance of the prognostic algorithm. $M$ cannot assume a negative value. In a loose sense, a connection can be identified between the metric $M$ and the Value of Information (VoI) metric \cite{Fauriat_2020,KAMARIOTIS2023109708} from Bayesian decision theory \cite{Raiffa_1961}.

    \begin{figure}
    \centering
    \includegraphics[width=0.85\linewidth]{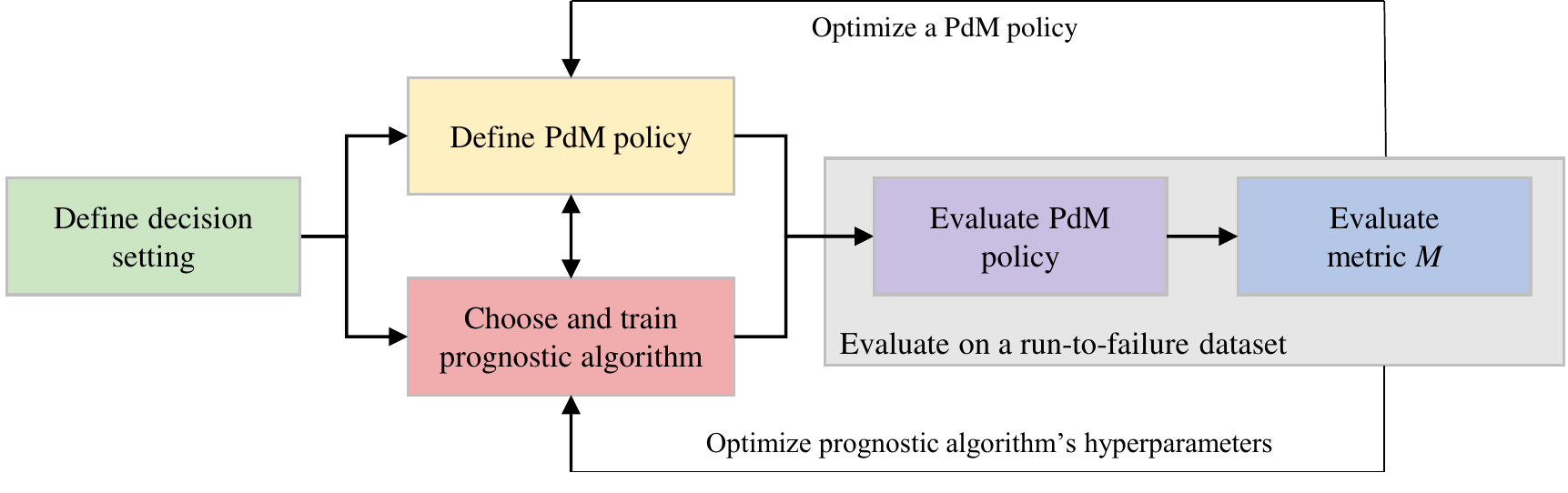}
    \caption{Summary of the proposed framework for a decision-oriented performance assessment and optimization of data-driven prognostic algorithms based on the proposed metric $M$.}
    \label{fig:synoptic_scheme}
    \end{figure}
 
    The estimate of \cref{eq:metric_hat} for the metric $M$ is subject to uncertainty. Assuming that the variance of $\hat{R}_\mathrm{perfect}$ is negligible, the variance of $\hat{M}$ is quantified as:
	\begin{equation}
        \text{Var}[\hat{M}] = \text{Var}\left[\dfrac{\hat{R}-\hat{R}_\mathrm{perfect}}{\hat{R}_\mathrm{perfect}}\right] = \text{Var}\left[\dfrac{\hat{R}}{\hat{R}_\mathrm{perfect}} - 1 \right] \approx \dfrac{1}{\hat{R}_\mathrm{perfect}^2}\cdot\text{Var}[\hat{R}]
        \label{eq:variance_M}
	\end{equation}
    Due to the assumption that the variance of $\hat{R}_\mathrm{perfect}$ is negligible, the covariance of $\hat{R}$ and $\hat{R}_\mathrm{perfect}$ is neglected.
	
    The decision-oriented metric $M$ can further serve as an objective function for optimizing PdM policies and for optimizing the training process of prognostics algorithms (e.g., hyperparameter tuning, see \cref{eq:hyperparameter-optimization}) directly with respect to subsequent PdM decision-making (see \cref{fig:synoptic_scheme}). This will be demonstrated in the numerical investigations of \cref{sec:Generic_example,sec:Case_study}.

    The metric $M$ is generally applicable in any given decision setting in conjunction with a PdM policy. The sections that follow specify the metric $M$ for two common decision settings.
    
    \subsection{PdM decision settings}
    In this paper, we consider and discuss two PdM decision settings that are typical for industrial assets. In the first basic setting (\cref{subsec:PdM_policy_1}), the only decision that has to be taken is when to replace a component. In the second setting (\cref{subsec:PdM_policy_2}), we additionally consider a decision on ordering and keeping a replacement component in the inventory. Both considered decision settings have the following characteristics:
	\begin{itemize}[noitemsep,nolistsep]
		\item We study single component problems.
		\item Continuous monitoring of a component is available.
        \item The monitoring data is employed to derive a probabilistic RUL prediction. 
		\item Inspections are not considered.
		\item Decisions on maintenance actions can only be made at discrete points in time, defined by $t_k = k\cdot\Delta T$, for a fixed time interval $\Delta T$ and integer $k=1,2,\dots,N$. The choice of $\Delta T$ is problem-dependent. For example, in the case of aeroengines, a value of $\Delta T=5-10$ flight cycles is deemed realistic based on the authors' expertise.
		\item Two types of replacement actions are considered:
		\begin{enumerate}[noitemsep,nolistsep]
			\itemsep0em 		
			\item Preventive replacement with cost $c_\mathrm{p}$.
			\item Corrective replacement with cost $c_\mathrm{c}$, which occurs upon component failure before a preventive replacement. Corrective replacement induces a larger cost than preventive replacement, which also accounts for longer downtime: $c_\mathrm{c}>c_\mathrm{p}$.
		\end{enumerate}
		\item A replacement is assumed to be a perfect replacement, bringing the component back to a pristine state. Replacement at $t$ leads to the end of one life-cycle of the component, and the component starts deteriorating anew. The stochastic deterioration process starting at time $t$ is a probabilistic copy of the process starting at time 0. These assumptions allow for use of renewal theory \cite{Tijms}. Within renewal theory, the time interval between two successive replacements defines a renewal cycle.
        \item Maintenance is a viable decision, i.e., it is possible to assume remediative action within the decision horizon.
	  \item It is assumed that failure is self-announcing.
        \item We do not include discounting of future costs.
	\end{itemize}
	
	\subsection{Predictive maintenance (PdM) planning for replacement}
	\label{subsec:PdM_policy_1}
	We first consider the simple dynamic PdM decision setting, in which one determines at each time step $t_k$ whether a component should be preventively replaced or not. The assumption here is that the new component is readily available when a preventive replacement is decided or a corrective replacement is imposed. 
	
    \subsubsection{Metric for prognostic performance evaluation with respect to PdM planning for replacement}
    \label{subsec:metric}
	
    Following a certain PdM policy for planning replacement (specific PdM policies are introduced in \cref{subsubsec:heur_policy_1,subsubsec:full_RUL_policy_1,subsubsec:novel_full_RUL_policy_1}), each $i$-th component life ends with a preventive replacement informed at time $T_\mathrm{R}^{(i)}$, or a corrective replacement in case of component failure at time $T_\mathrm{F}^{(i)}$. A preventive replacement can only be performed at discrete points in time, i.e., $T_\mathrm{R}^{(i)}$ lies in the set $\{t_k=k\cdot \Delta T, k=1,2,\dots\}$. A corrective replacement is performed immediately upon failure at $T_\mathrm{F}^{(i)}$. The (non-discounted) cost of the replacement action for the $i$-th component is
	\begin{equation}
		C_{\mathrm{rep}}^{(i)} = \begin{cases}
			c_\mathrm{p}, & \text{if $T_\mathrm{R}^{(i)}<T_\mathrm{F}^{(i)}$}\\
			c_\mathrm{c}, & \text{else.} 
		\end{cases}
    \label{eq:cost_replacement}
	\end{equation}
	Replacement (preventive or corrective) leads to the end of one life-cycle of a component. The length of the life-cycle of the $i$-th component is thus $T_{\mathrm{lc}}^{(i)}= \text{min}[T_\mathrm{R}^{(i)},T_\mathrm{F}^{(i)}]$. In this setting, the cost of maintenance for the $i$-th component is equal to the cost of the replacement action, i.e., $C_\mathrm{m}^{(i)}=C_{\mathrm{rep}}^{(i)}$.

    Within the current decision setting, the perfect PdM policy would not lead to any corrective replacement, as this is more costly, or any early preventive replacement, as this leads to shortening of the component life-cycle. Thus, the perfect PdM policy is a preventive replacement with cost $C_\mathrm{m,perfect}^{(i)}=c_\mathrm{p}$ at the optimal time step $t_k = k\cdot \Delta T$ directly before $T_\mathrm{F}^{(i)}$ \footnote{In extreme cases, where $c_\mathrm{c}$ is very close to $c_\mathrm{p}$, and where $\Delta T$ is small enough, the optimal action may be to allow the component to fail at $T_\mathrm{F}^{(i)}$ with a cost $C_\mathrm{m,perfect}^{(i)}=c_\mathrm{c}$.}. This is denoted by $T_\mathrm{R,perfect}^{(i)}$. The long-run expected maintenance cost per unit time of the perfect PdM policy is evaluated via \cref{eq:ratio_perfect_PdM}.

    Eventually, the decision-oriented metric proposed in \cref{eq:metric}, in particular in conjunction with a PdM policy for replacement, is estimated as:
    \begin{equation}
    \hat{M} = \dfrac{\dfrac{\frac{1}{n}\sum_{i=1}^n C_{\mathrm{rep}}^{(i)}}{\frac{1}{n}\sum_{i=1}^n T_{\mathrm{lc}}^{(i)}} - \dfrac{c_\mathrm{p}}{\frac{1}{n}\sum_{i=1}^n T_{\mathrm{R,perfect}}^{(i)}}}{\dfrac{c_\mathrm{p}}{\frac{1}{n}\sum_{i=1}^n T_{\mathrm{R,perfect}}^{(i)}}}.
    \label{eq:metric1}
    \end{equation}

    \subsubsection{PdM policy 1: simple heuristic PdM policy for preventive replacement}
    \label{subsubsec:heur_policy_1}
	
    The first PdM policy that we consider is a simple heuristic policy, similar to \cite{Nguyen_2019}. Heuristic policies employ simple and intuitive decision rules that are easily understood by engineers and operators \cite{Luque_2019}. Specifically, at each time step $t_k = k \cdot \Delta T$, the policy determines the action $a_{\mathrm{rep},k}$ to take as: 
    \begin{equation}
		a_{\mathrm{rep},k} = \begin{cases}
			\text{DN}, & \text{if $\text{Pr}(RUL_{\mathrm{pred},k}\leq\Delta T)<p_\textrm{thres}$}\\
			\text{PR} & \text{else,} 
		\end{cases}
	\label{eq:heuristic1}
    \end{equation}
    where $\text{DN}$ denotes the do nothing action, $\text{PR}$ denotes the preventive replacement action, $RUL_{\mathrm{pred},k}$ is the RUL prediction estimated at time $t_k$ from the employed prognostic algorithm, and $p_\textrm{thres}$ is a variable heuristic threshold. In \cite{Nguyen_2019}, a value of $p_\textrm{thres}=c_\mathrm{p}/c_\mathrm{c}$ has been used. The $\text{PR}$ action is associated with a cost $c_\mathrm{p}$, while the $\text{DN}$ action entails the predicted risk of component failure within the next time step, $\text{Pr}(RUL_{\mathrm{pred},k}\leq\Delta T)\cdot c_\mathrm{c}$. The reasoning behind $p_\textrm{thres}=c_\mathrm{p}/c_\mathrm{c}$ is that the PR action is performed at time $t_k$ only when the associated cost is smaller than the predicted risk of component failure in the next time step. This is a simplification, as it does not account for the future time steps; after a replacement, one has a new component with - on average - lower maintenance costs. Hence the choice $p_\textrm{thres}=c_\mathrm{p}/c_\mathrm{c}$ leads to suboptimal decisions. An improvement can be reached by optimizing the heuristic threshold $p_\mathrm{thres}$, as we demonstrate in the numerical investigations of \cref{sec:Generic_example,sec:Case_study}.
    
    The simple heuristic PdM policy requires the predicted probability of RUL exceedance within the next decision time step, $\text{Pr}(RUL_{\mathrm{pred},k}\leq\Delta T)$, as sole input from the prognostics, in order to evaluate the DN versus PR decision at each time step. Different prognostic algorithms operate in distinct manners, and thus use different methods for deriving this probability. This can be specified as the probability of the $RUL_{\mathrm{pred},k}$ belonging to a certain class (corresponding to $RUL_{\mathrm{pred},k}\leq\Delta T$) in the case of a prognostic classifier \cite{Nguyen_2019}. For the case of prognostic regression models, uncertainty quantification in the RUL predictions is a prerequisite for obtaining this probability.
	
    Eventually, this heuristic PdM policy informs the replacement decision for each $i$-th component, leading to $C_\mathrm{rep}^{(i)}$ and $T_\mathrm{lc}^{(i)}$. The long-run expected maintenance cost per unit time associated with this heuristic PdM policy can then be evaluated via \cref{eq:ratio_MCS}.
	
	\subsubsection{PdM policy 2: PdM policy for preventive replacement on the basis of the full RUL distribution.}
	\label{subsubsec:full_RUL_policy_1}
	
    \begin{figure}
    \centering
    \includegraphics[width=0.46\linewidth]{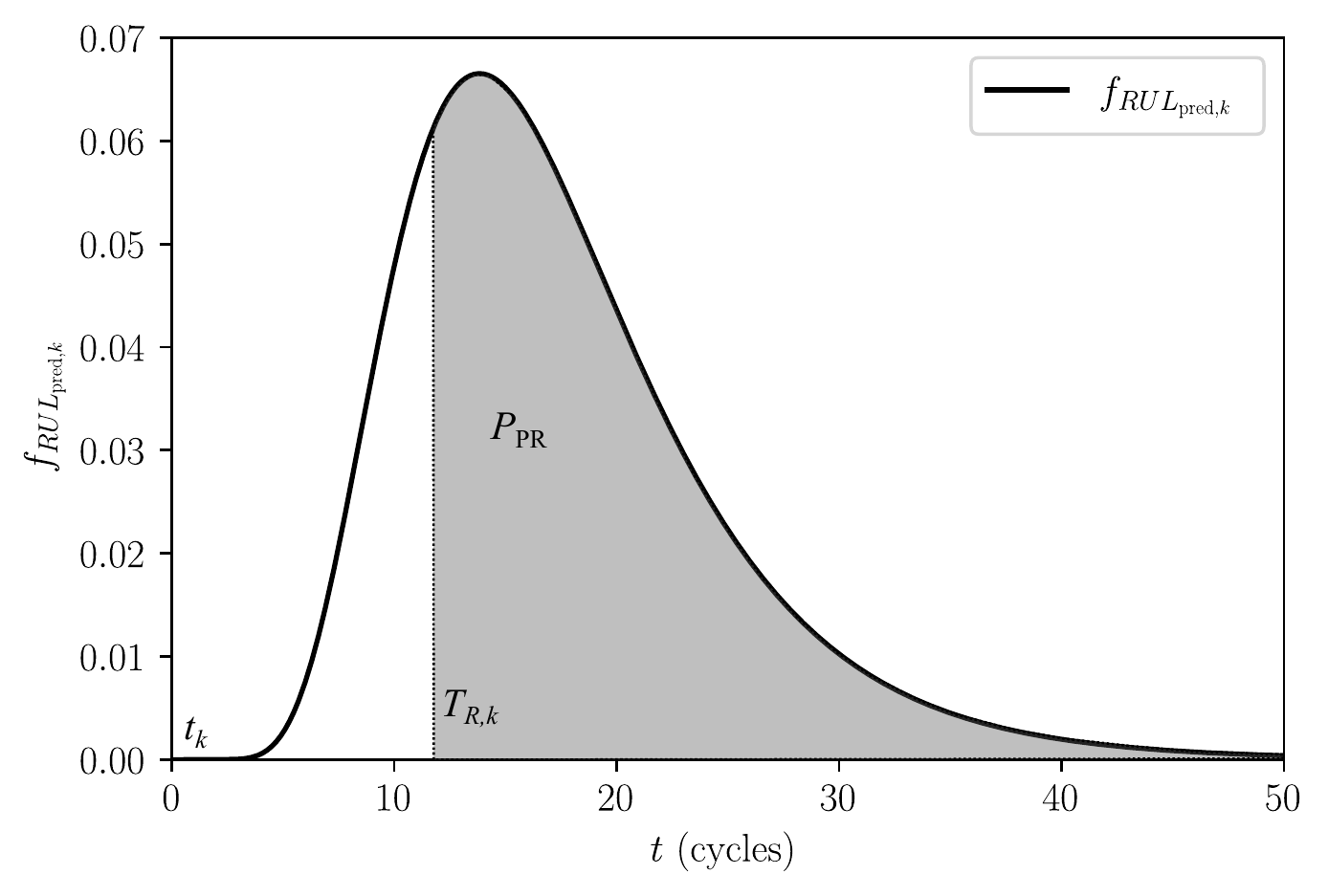}
    \caption{Optimization of $T_{\mathrm{R},k}$. $P_\mathrm{PR}$ corresponds to the probability of a preventive replacement for a fixed $T_{\mathrm{R},k}$, which appear in the objective function of \cref{eq:int_obj_1}}
    \label{fig:RUL_policy_P_PR}
    \end{figure}

    We consider a second PdM policy for preventive replacement, which determines at each time step $t_k$ the action $a_{\mathrm{rep},k}$ to take as:
    \begin{equation}
		a_{\mathrm{rep},k} = \begin{cases}
			\text{PR}, & \text{if $t_k+\Delta T\geq T^*_{\mathrm{R},k}$}\\
			\text{DN} & \text{else,} 
		\end{cases}
	\label{eq:decision_2}
    \end{equation}
    where $T^*_{\mathrm{R},k}$ is the optimal time to replacement found through solving an optimization problem at each time step $t_k$. A preventive replacement is thus decided when $T^*_{\mathrm{R},k}$ is smaller or equal to $t_k + \Delta T$, where $\Delta T$ defines the time interval until the next decision.

    The most commonly employed objective function for finding the optimal $T^*_{\mathrm{R},k}$ in RUL-based PdM \cite{Fauriat_2020, YAN2022109053, Zeng_2022} is presented in this section. It can be employed when the full distribution of the RUL prediction at each time step $t_k$ is available, denoted by $f_{RUL_{\mathrm{pred},k}}(t)$. The distribution of the predicted time to failure ($T_F$) at time step $t_k$ is $f_{T_{F\mathrm{pred},k}}(t)=f_{RUL_{\mathrm{pred},k}}(t-t_k)$, and is bounded below at $t_k$. The objective function for the optimization problem at each time step $t_k$ is:
    \begin{equation}
        f(T_{\mathrm{R},k}) = \frac{\text{E}[C_\mathrm{rep}(T_{\mathrm{R},k})]}{\text{E}[T_\mathrm{lc}(T_{\mathrm{R},k})]} =  \frac{P_\mathrm{PR}\cdot c_\mathrm{p} + (1-P_\mathrm{PR})\cdot c_\mathrm{c}}{P_\mathrm{PR}\cdot(T_{\mathrm{R},k}) + \int\limits_{t}^{T_{\mathrm{R},k}} t\, f_{RUL_{\mathrm{pred},k}}(t-t_k)\,\mathrm{d} t},
	\label{eq:int_obj_1}
    \end{equation}
    where: 
    \begin{equation}
    P_\mathrm{PR} = \int\limits_{T_{\mathrm{R},k}}^{\infty} f_{RUL_{\mathrm{pred},k}}(t-t_k)\,\mathrm{d} t
    \label{eq:PPR}
    \end{equation}
    denotes the probability that the component will be preventively replaced at $T_{\mathrm{R},k}$, whereas $(1-P_\mathrm{PR})$ is the probability that the component will fail before $T_{\mathrm{R},k}$ with an induced cost $c_\mathrm{c}$. These probabilities are graphically represented for a fixed $T_{\mathrm{R},k}$ in \cref{fig:RUL_policy_P_PR}.
 
    The objective function in \cref{eq:int_obj_1} makes use of renewal theory and evaluates the long-run expected maintenance cost per unit time, in order to determine the optimal time for preventive replacement for a single component. The underlying assumption when using this objective function for optimizing replacement for a single component is that the predicted distribution of the time to failure of the component, $f_{T_{F\mathrm{pred},k}}(t)$, corresponds to the underlying distribution of the time to failure of the whole population of components. This (incorrect) assumption has not been clarified in the literature that uses this policy.
	
	Eventually, this PdM policy optimizes the replacement decisions for each $i$-th component leading to $C_\mathrm{rep}^{(i)}$ and $T_\mathrm{lc}^{(i)}$. The long-run expected maintenance cost per unit time of this PdM policy can then be quantified via \cref{eq:ratio_MCS}.
	
	\subsubsection{PdM policy 3: modified PdM policy for preventive replacement on the basis of the full RUL distribution}
	\label{subsubsec:novel_full_RUL_policy_1}
	In this section we propose a modification of the objective function defined in \cref{eq:int_obj_1} to overcome the implicit assumption pointed out above, i.e., we no longer assume that all future components have the same lifetime distribution as the one implied by the RUL prediction of the current component. Instead, we approximate the long-run expected maintenance cost per unit time of all future components, $\frac{\text{E}_{\bar{T}_\mathrm{F}}[C_\mathrm{rep}]}{\text{E}_{\bar{T}_\mathrm{F}}[T_\mathrm{lc}]}$, and utilize it to formulate the objective function for the time to replace the current component. This objective function is the sum of the cost associated with maintenance of the current component and the ``opportunity'' loss associated with replacing the component too early\footnote{Please note that the objective functions in \cref{eq:int_obj_1,eq:novel_int_obj_1} have different units, as \cref{eq:int_obj_1} quantifies a cost per unit time, while \cref{eq:novel_int_obj_1} quantifies a cost.}. This opportunity loss is equal to the expected value of the lifetime of the current component beyond the time of replacement $T_{\mathrm{R},k}$ multiplied with $\frac{\text{E}_{\bar{T}_\mathrm{F}}[C_\mathrm{rep}]}{\text{E}_{\bar{T}_\mathrm{F}}[T_\mathrm{lc}]}$. The resulting objective function for finding the optimal $T^*_{\mathrm{R},k}$ at each time step $t_k$ is:
	\begin{equation}
	    f(T_{\mathrm{R},k}) = P_\mathrm{PR}\cdot c_\mathrm{p} + (1-P_\mathrm{PR})\cdot c_\mathrm{c} + \frac{\text{E}_{\bar{T}_\mathrm{F}}[C_\mathrm{rep}]}{\text{E}_{\bar{T}_\mathrm{F}}[T_\mathrm{lc}]}\,\int\limits_{T_{\mathrm{R},k}}^{\infty} (t-T_{\mathrm{R},k}) \cdot f_{RUL_{\mathrm{pred},k}}(t-t_k)\,\mathrm{d} t, 
	\label{eq:novel_int_obj_1}
	\end{equation}
	where $P_\mathrm{PR}$ is defined in \cref{eq:PPR}. The first two terms of \cref{eq:novel_int_obj_1} (which correspond to the numerator of the objective function in \cref{eq:int_obj_1}) are the expected replacement cost of the component. The integral in the last term of \cref{eq:novel_int_obj_1} is the expected lifetime of the component beyond $T_{\mathrm{R},k}$.

To approximate the long-run expected maintenance cost per unit time of all future components, $\frac{\text{E}_{\bar{T}_\mathrm{F}}[C_\mathrm{rep}]}{\text{E}_{\bar{T}_\mathrm{F}}[T_\mathrm{lc}]}$, we make use of the estimated distribution $f_{\bar{T}_\mathrm{F}}$ of the time to failure of the population of components. 
We consider the following assumptions to approximate $\frac{\text{E}_{\bar{T}_\mathrm{F}}[C_\mathrm{rep}]}{\text{E}_{\bar{T}_\mathrm{F}}[T_\mathrm{lc}]}$:
    \begin{enumerate}[noitemsep,nolistsep]
        \item For an assumed case without monitoring, renewal theory can be used to find the optimal time for preventive replacement with respect to $f_{\bar{T}_\mathrm{F}}$. Then $\frac{\text{E}_{\bar{T}_\mathrm{F}}[C_\mathrm{rep}]}{\text{E}_{\bar{T}_\mathrm{F}}[T_\mathrm{lc}]}$ can be set equal to the corresponding optimal value of the long-run expected maintenance cost per unit time. This choice delivers an upper bound to the value of this term, causing early preventive replacements to be penalized more, and consequently delivers a less conservative PdM policy.
        \item For an assumed ``perfect" monitoring case, replacement of every component will be a preventive one, and the expected life-cycle length of the population of components will be equal to the mean $\mu_{\bar{T_F}}$ of the distribution $f_{\bar{T}_\mathrm{F}}$. Therefore, one can set $\frac{\text{E}_{\bar{T}_\mathrm{F}}[C_\mathrm{rep}]}{\text{E}_{\bar{T}_\mathrm{F}}[T_\mathrm{lc}]} = \frac{c_\mathrm{p}}{\mu_{\bar{T_\mathrm{F}}}}$. This choice yields a lower bound to the value of this term, leading to a more conservative PdM policy.
        \item A value for $\frac{\text{E}_{\bar{T}_\mathrm{F}}[C_\mathrm{rep}]}{\text{E}_{\bar{T}_\mathrm{F}}[T_\mathrm{lc}]}$ between the upper bound of option 1 and the lower bound of option 2 can be chosen, e.g., the average of these bounds. 
    \end{enumerate}
	
	\subsection{Predictive maintenance (PdM) planning for component ordering and replacement}
	\label{subsec:PdM_policy_2}
	
    In the first decision setting of \cref{subsec:PdM_policy_1}, it is assumed that the new component will always be available for replacement. In this section, we consider a second decision setting, which includes ordering and replacement decisions. A deterministic lead time $L$ is assumed from the time of component ordering $T_\mathrm{order}$ to the time of component delivery. We implicitly assume that $L$ is a multiple of $\Delta T$.
	
	\subsubsection{Metric for prognostic performance evaluation with respect to PdM planning for ordering \& replacement}

    Following a certain PdM policy for component ordering and replacement (one specific policy is introduced in \cref{subsubsec:heur_policy_2}), different costs will be induced. The cost of the replacement action for the $i$-th component is given by \cref{eq:cost_replacement}.
	
	The cost related to a late ordering of a component for replacement (preventive or corrective) is:
	\begin{equation}
		C_\mathrm{delay}^{(i)} = \text{max}\big(T_\mathrm{order}^{(i)} + L - T_\mathrm{lc}^{(i)}, 0\big) \cdot c_\mathrm{unav},
		\label{eq:cost_delay}
	\end{equation}
	where $c_\mathrm{unav}$ is the system unavailability cost per unit time, related to necessary operation shutdown from $T_\mathrm{lc}^{(i)}$ until the time of component arrival, upon which a replacement can be performed.
	
	The cost related to an early ordering of a component, i.e., the holding inventory cost, is:
	\begin{equation}
		C_\mathrm{stock}^{(i)} = \text{max}\big(T_\mathrm{lc}^{(i)} - (T_\mathrm{order}^{(i)}+L), 0\big)\cdot c_\mathrm{inv},
		\label{eq:cost_stock}
	\end{equation}
	where $c_\mathrm{inv}$ is the holding inventory cost per unit time for a component.
	
	The total maintenance cost, excluding the cost of the new component, is:
	\begin{equation}
		C_\mathrm{m}^{(i)} = C_\mathrm{rep}^{(i)} + C_\mathrm{delay}^{(i)} + C_\mathrm{stock}^{(i)}
		\label{eq:cost_total}
	\end{equation}
 
    The long-run expected maintenance cost per unit time that is achieved with a PdM policy is evaluated by applying the PdM policy on $n$ independent components via \cref{eq:ratio_MCS}.

    In this second decision setting, the perfect PdM policy would replace the $i$-th component at $T_\mathrm{R,perfect}^{(i)}$, which is the time step $k\cdot \Delta T$ directly before $T_\mathrm{F}^{(i)}$, and order a component at $T_\mathrm{order,perfect}^{(i)} = T_\mathrm{R,perfect}^{(i)} - L$. Therefore, the costs induced when applying the perfect PdM policy to the $i$-th component are $C_\mathrm{stock, perfect}^{(i)}=0$, $C_\mathrm{delay, perfect}^{(i)}=0$, and $C_\mathrm{rep, perfect}^{(i)}=c_\mathrm{p}$, resulting in $C^{(i)}_\mathrm{m, perfect}=c_\mathrm{p}$. 

    Eventually, the decision-oriented metric for component ordering and replacement is estimated as:
    \begin{equation}
    \hat{M} = \dfrac{\dfrac{\frac{1}{n}\sum_{i=1}^n (C_\mathrm{rep}^{(i)} + C_\mathrm{delay}^{(i)} + C_\mathrm{stock}^{(i)})}{\frac{1}{n}\sum_{i=1}^n T_{\mathrm{lc}}^{(i)}} - \dfrac{c_\mathrm{p}}{\frac{1}{n}\sum_{i=1}^n T_{\mathrm{R,perfect}}^{(i)}}}{\dfrac{c_\mathrm{p}}{\frac{1}{n}\sum_{i=1}^n T_{\mathrm{R,perfect}}^{(i)}}}.
    \label{eq:metric2}
    \end{equation}

    \subsubsection{Simple heuristic PdM policy for component ordering and preventive replacement}
    \label{subsubsec:heur_policy_2}
	
	At each time step $t_k$ (as long as no replacement component has been ordered previously), the policy first determines based on the prognostics input whether a replacement component should be ordered (O), or not (NO). The considered simple policy determines the action to take as:
	\begin{equation}
		a_{\mathrm{order},k} = \begin{cases}
			\text{O}, & \text{if $\text{Pr}(RUL_{\mathrm{pred},k}\leq w+\Delta T)\geq p_\mathrm{thres}^\mathrm{order}$}\\
			\text{NO} & \text{else,} 
		\end{cases}
		\label{eq:order_heuristic}
	\end{equation}
	where $w= \left\lceil\dfrac{L}{\Delta T}\right\rceil\cdot \Delta T$ is the ordering lead time adjusted for the discrete time steps. $p_\mathrm{thres}^\mathrm{order}$ is a variable heuristic threshold. 
 
	The reasoning behind the condition of \cref{eq:order_heuristic} is the following: a lead time $L$ is required for the component to become available upon ordering it. Therefore, if a component is ordered at time step $t_k$, the earliest future decision time at which the component will be available for replacement is $t_k + w$. The simple policy assumes that the critical threshold for the O-NO decision is based on the predicted probability that a preventive replacement will be necessary at time $t_k + (w+\Delta T)$. Once a component has been ordered, the O-NO decision is no longer relevant until a replacement action is performed.
	
	At each time step $t_k$, the policy further determines whether the component is preventively replaced (PR) or nothing is done (DN). This applies independent of whether or not a new component is in stock, which, as discussed in \cref{subsec:CMAPSS_second_setting}, is not an optimal choice. Similar to \cref{subsubsec:heur_policy_1}, the policy determines the action to take $a_{\textrm{rep},k}$ as:
	\begin{equation}
		a_{\textrm{rep},k} = \begin{cases}
			\text{DN}, & \text{if $\text{Pr}(RUL_{\mathrm{pred},k}\leq\Delta T)<p^\mathrm{rep}_\mathrm{thres}$}\\
			\text{PR} & \text{else.} 
		\end{cases}
		\label{eq:replacement_heuristic_2}
	\end{equation}

    A value of $p^\mathrm{rep}_\mathrm{thres}= p^\mathrm{order}_\mathrm{thres}=c_\mathrm{p}/c_\mathrm{c}$ has been used in literature \cite{Nguyen_2019}. The values of the two heuristic thresholds may be optimized, leading to an improvement of this heuristic policy, as we demonstrate in \cref{subsec:CMAPSS_second_setting}.
	
    The advantage of this heuristic PdM policy lies in its simplicity and universal applicability -- it can be applied as long as the RUL prediction provides the probabilities in \cref{eq:order_heuristic,eq:replacement_heuristic_2}. Other, more complex and possibly more optimal heuristic policies, or even algorithms such as partially observable Markov decision processes (POMDP) \cite{PAPAKONSTANTINOU2014202,Arcieri_2022}, can be investigated, but we leave this for future work.

	\section{Numerical investigations on a virtual RUL simulator}
        \label{sec:Generic_example}
        
    This section investigates the proposed metric and the PdM policies presented in \cref{subsec:PdM_policy_1} in a hypothetical setup. For this purpose, a virtual RUL simulator serves as a test-bed, which enables the assessment and evaluation of PdM policies for varying data availability. The aim of this section is to investigate and quantitatively assess the performance and optimality of the three PdM policies for replacement and their effect on metric $M$, and eventually propose a set of directives towards optimizing decision heuristics.

    \subsection{Virtual RUL simulator}
 
    In this section we introduce a virtual RUL simulator, i.e., we establish a model with which to generate $RUL_{\textrm{pred},k}$ distributions over time, conditional on given underlying ``true" realizations of the failure time of a component, emulating uncertain RUL predictions provided by a prognostic algorithm in a realistic setting. 
  
    We assume that the uncertain time to failure (expressed in cycles to failure) of a hypothetical population of mechanical components follows a normal distribution, specifically $T_\mathrm{F}\sim N(\mu=225, \sigma = 40)$, where $N$ denotes the Gaussian probability density function (PDF) with mean $\mu$ and standard deviation $\sigma$. We draw samples from this distribution, each representing one underlying ``true" realization of a component's failure time.
	
	We define a PdM planning problem, wherein it is assumed that the maintenance actions can only be performed at discrete points in time $t_k$, defined by $t_k=k\cdot\Delta T$, for fixed $\Delta T=10$ cycles and $k=1,\dots,N$. The predicted time to failure, yielded as an output of a prognostic algorithm at time step $k$, is denoted by $T_{\textrm{F,pred},k}$. \cref{eq:pred_ln_error} defines in logarithmic scale the modeled discrepancy between the prognostic RUL prediction at time step $k$, $RUL_{\mathrm{pred},k}=T_{\textrm{F,pred},k}-t_k$, and the underlying ``true" RUL value at time step $k$, $RUL_k=T_\mathrm{F} -t_k$:
 	\begin{equation}
	\text{ln}\left(RUL_{\textrm{pred},k}\right) = \text{ln}(RUL_k) + \text{ln}(\epsilon_k),
	\label{eq:pred_ln_error}
	\end{equation} 
 where $\epsilon_k$ is the prognostic prediction error. We assume the following probabilistic model for the random vector containing the logarithm of the prediction errors over time
	\begin{equation}
	[\text{ln}(\epsilon_1),\dots, \text{ln}(\epsilon_n)] \sim \text{MVN}\left(\mathbf{0}, \mathbf{\Sigma}\right), 
	\label{eq:MVN}
	\end{equation}  
	where $\text{MVN}()$ denotes the multivariate normal distribution with mean vector $\mathbf{0}$ and covariance matrix $\mathbf{\Sigma}$, which is constructed as
	\begin{equation}
	\mathbf{\Sigma} = \mathbf{D}\cdot\mathbf{R}\cdot\mathbf{D}, 
	\end{equation}  
	where $\mathbf{D}$ is a diagonal matrix containing the standard deviation of the prediction errors $\sigma_{\text{ln}(\epsilon_k)}$ in the diagonal, and $\mathbf{R}$ is a correlation coefficient matrix. The prediction errors over time are assumed to be correlated according to an exponential correlation model with correlation length $l$:
	\begin{equation}
	\mathbf{R} = [\rho_{ij}], \quad \text{where} \quad \rho_{ij}=\text{exp}\left(-\dfrac{|t_{i} - t_{j}|}{l}\right)
	\end{equation}  
	
    \noindent With this model, the distributions $f_{RUL_{\textrm{pred},k}}$ are obtained as follows:
	\begin{itemize}[noitemsep,nolistsep]
	    \item Sample the mean values $\mu_{\text{ln}\left(RUL_{\textrm{pred},k}\right)}$ of the different $\text{ln}\left(RUL_{\textrm{pred},k}\right)$ predictions via \cref{eq:sample_mean_RUL} by drawing a sample $[\text{ln}(\epsilon_1^{(i)}),\dots, \text{ln}(\epsilon_n^{(i)})]$ from the MVN distribution of \cref{eq:MVN}.
	    \begin{equation}
	    \mu^{(i)}_{\text{ln}(RUL_{\textrm{pred},k})} = \text{ln}(RUL^{(i)}_k) + \text{ln}(\epsilon_k^{(i)})
	    \label{eq:sample_mean_RUL}
	    \end{equation}
	    \item $\text{ln}\left(RUL_{\textrm{pred},k}\right)$ then follows the normal distribution with mean equal to the value sampled in \cref{eq:sample_mean_RUL} and standard deviation $\sigma_{\text{ln}(\epsilon_k)}$, i.e., $\text{ln}\left(RUL_{\textrm{pred},k}\right)\sim N\left(\mu = \mu^{(i)}_{\text{ln}\left(RUL_{\textrm{pred},k}\right)}, \sigma = \sigma_{\text{ln}(\epsilon_k)}\right)$.  
	\end{itemize}

\begin{figure}
	\centering
		\includegraphics[width=.50\linewidth]{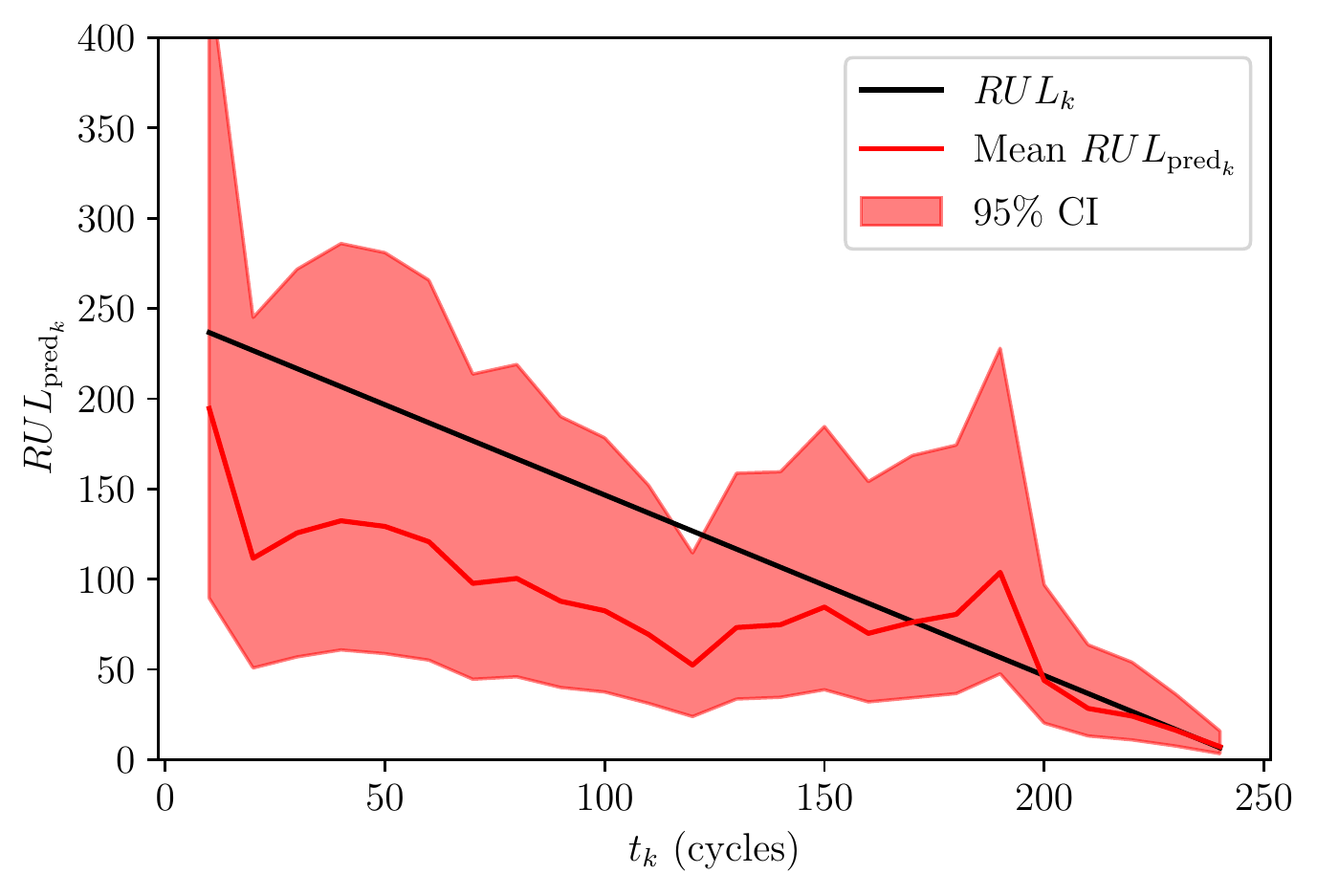}
	\caption{Demonstration of the virtual RUL simulator. Generated $RUL_{\textrm{pred},k}$ distributions conditional on an underlying true realization $T_\mathrm{F}^{(i)}=247$ cycles.}
	\label{fig:RUL_pred_k}
	\end{figure}

	For an underlying ``true" realization $T_\textrm{F}^{(i)}=247$, a sampled realization of means of the different $RUL_{\textrm{pred},k}$ distributions over time and the 95\% credible intervals (CI), are plotted in \cref{fig:RUL_pred_k}. For generating these distributions, we have assumed $\sigma_{\text{ln}(\epsilon_k)}=0.4$ and correlation length $l=50$ cycles in the definition of the prediction errors. 
	
	The useful feature of this virtual RUL simulator is that one can generate a large number of virtual run-to-failure experiments via a Monte Carlo simulation. Each sample consists of a $T_\textrm{F}^{(i)}$ sample and associated $RUL_{\textrm{pred},k}$ distributions, assumed to be obtained via fusion of monitoring data with a prognostic algorithm, simplistically simulating realistic PHM settings. Using these samples, the ratio in \cref{eq:ratio_MCS} is evaluated for a PdM policy and subsequently the metric $M$ corresponding to a decision setting is estimated via \cref{eq:metric_hat}. Introducing this virtual RUL simulator allows the generation of a large number of samples, with which we investigate the performance of the three  PdM policies for replacement of \cref{subsec:PdM_policy_1} and the uncertainty in estimating the metric $M$ via \cref{eq:metric1} in function of the number of available run-to-failure experiments.
	
	\subsection{First decision setting: PdM planning for replacement}
	\label{subsec:generic_1}	
\begin{figure}
	\centering
	\begin{subfigure}{.495\textwidth}
	\includegraphics[width=1.\linewidth]{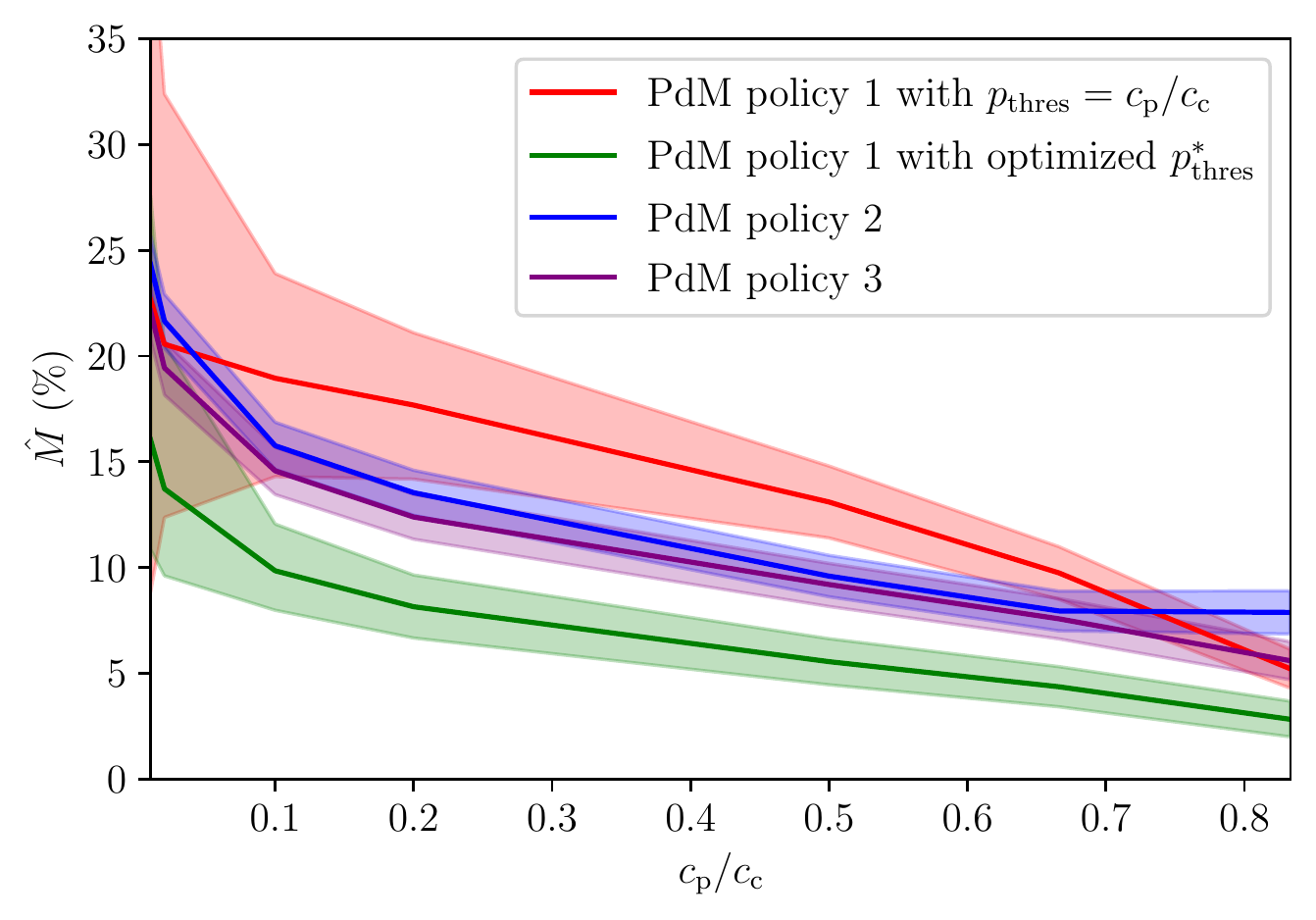}
	\caption{Estimation of the metric $M$ via \cref{eq:metric_hat}, and the associated uncertainty (95\% CIs), in combination with the PdM policies of \cref{subsec:PdM_policy_1}, for different $c_\mathrm{p}/c_\mathrm{c}$ ratios.}
        \label{subfig:3a}
	\end{subfigure}
	\begin{subfigure}{.495\textwidth}
	\includegraphics[width=1.\linewidth]{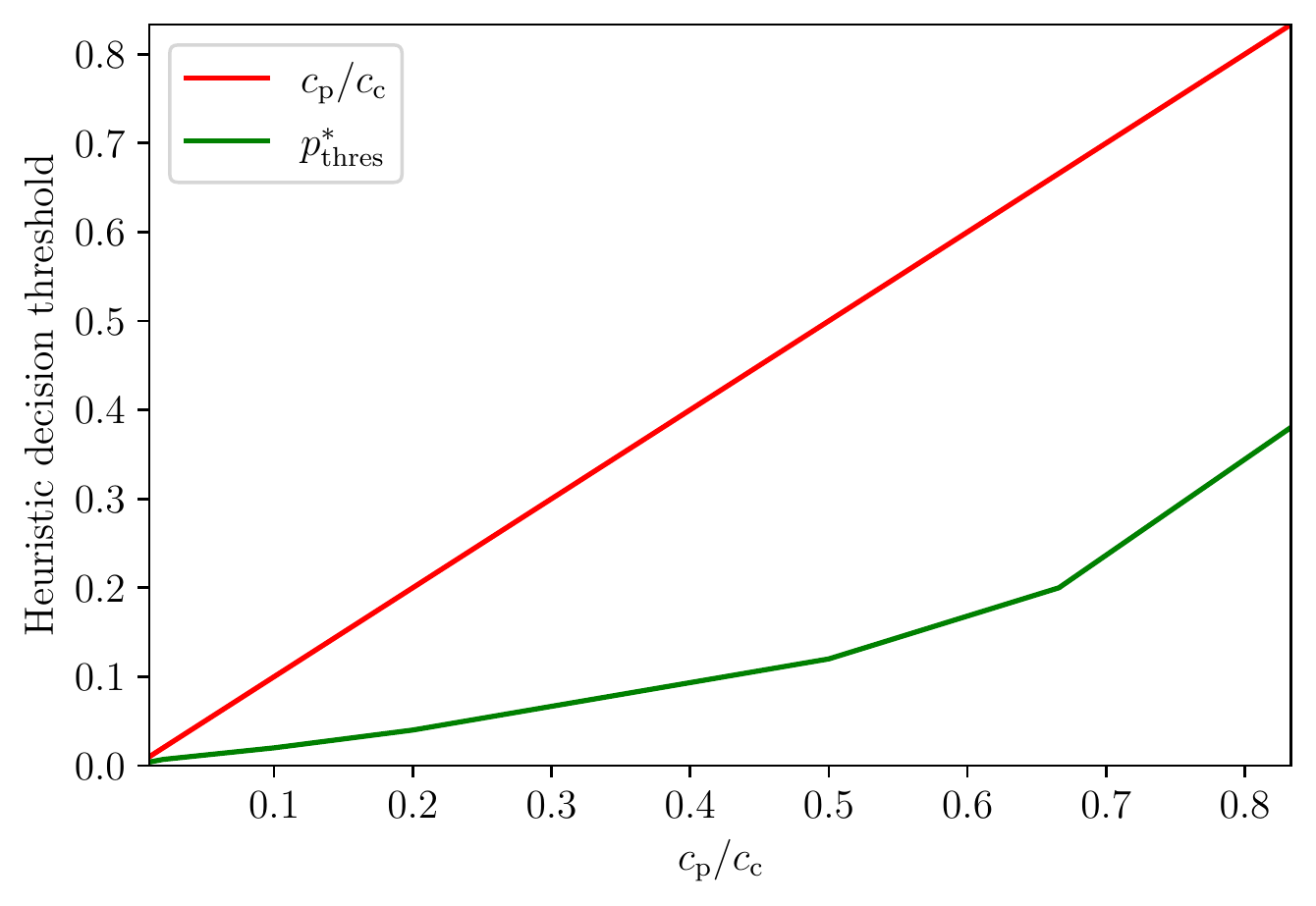}
	\caption{The probability threshold that is used as a heuristic decision threshold in the PdM policy 1. The green line plots the optimal threshold value found for different $c_\mathrm{p}/c_\mathrm{c}$ ratios.}
         \label{subfig:3b}
	\end{subfigure}
	\caption{Results from virtual RUL simulator with $n=2\cdot10^3$ sampled components for $\sigma_{\text{ln}(\epsilon_k)}=0.4$.}
	\label{fig:error_M_generic_1}
	\end{figure}

    To investigate the first decision setting, we generate $n = 2\cdot10^3$ samples of $T_\textrm{F}\sim N(\mu=225, \sigma = 40)$ and corresponding $RUL_{\textrm{pred},k}$ distributions (generated for $\sigma_{\text{ln}(\epsilon_k)}=0.4$). These correspond to $n = 2\cdot10^3$ hypothetical components, for which run-to-failure monitoring data, as well as predicted $RUL_{\textrm{pred},k}$ distributions provided by means of a prognostic algorithm, are assumed available. We employ all three PdM policies presented in \cref{subsec:PdM_policy_1}, and for each component we compute the $C^{(i)}_\textrm{rep}$ and $T^{(i)}_\textrm{lc}$ for each of the three policies, for different cost ratios $c_\textrm{p}/c_\textrm{c}$. Subsequently, for each policy we evaluate the long-run expected maintenance cost per unit time via \cref{eq:ratio_MCS}, and estimate the metric $M$ via \cref{eq:metric1}. In \cref{subfig:3a} we plot the value of $\hat{M}$ and the associated uncertainty, represented with 95\% CIs, in function of the cost ratio. The red line corresponds to the values of $\hat{M}$ when using the heuristic PdM policy 1 with $p_\mathrm{thres}=c_\mathrm{p}/c_\mathrm{c}$. The blue line corresponds to use of the PdM policy 2, whereas the purple line corresponds to the PdM policy 3, where option 1 has been chosen for the value of $\frac{\text{E}_{\bar{T}_\mathrm{F}}[C_\mathrm{rep}]}{\text{E}_{\bar{T}_\mathrm{F}}[T_\mathrm{lc}]}$. Finally, the green line corresponds to the optimized heuristic PdM policy 1, wherein the heuristic threshold has been optimized. The optimal threshold $p^*_\mathrm{thres}$ is found as the argument that minimizes the metric $M$, when estimating it using \cref{eq:metric1}, with $C^{(i)}_\textrm{rep}$ and $T^{(i)}_\textrm{lc}$ found by applying the heuristic PdM policy 1 on each of the $n=2\cdot10^3$ components. The values that  $p_\textrm{thres}^*$ assumes for the different cost ratios are plotted in \cref{subfig:3b}. It is noted that for all cost ratios, the optimal value of $p_\textrm{thres}^*$ is smaller than $c_\textrm{p}/c_\textrm{c}$.
	
	Various conclusions can be drawn from the results of this numerical investigation. For most cost ratios, the PdM policies 2 and 3, which operate on the basis of the availability of the full RUL distribution, lead to better performance than the heuristic PdM policy 1 and to reduced uncertainty. For the relatively large level of prognostic uncertainty considered here ($\sigma_{\text{ln}(\epsilon_k)}=0.4$), the PdM policies 2 and 3 tend to be more conservative than the heuristic PdM policy, leading to earlier preventive replacements (for some components significantly earlier) in order to reduce the risk of corrective failure, but also reducing the life-cycle of the components, especially for low values of $c_\textrm{p}/c_\textrm{c}$. The heuristic PdM policy 1 instead informs later preventive replacements, which is favorable for many components, but at the same time leads to corrective replacements for some components, even for cases when $c_\mathrm{c}$ is significantly large. The latter is the reason for its seemingly worse performance and its associated increased uncertainty in the evaluation of the metric $M$. Naturally, as the ratio $c_\textrm{p}/c_\textrm{c}$ increases, corrective replacements become less critical, and all three policies do lead to some corrective replacements. It is noted that even with a relatively large number of samples, the evaluation of all considered PdM policies with \cref{eq:ratio_MCS}, and the estimation of $M$ with \cref{eq:metric_hat}, seem to entail non-negligible uncertainty.
	
    The PdM policy 3 that we propose in \cref{subsubsec:novel_full_RUL_policy_1} proves somewhat less conservative, and thus delivers better results than the PdM policy 2, which is the one most widely used in literature. Optimizing the heuristic PdM policy leads to a policy that delivers the best performance among all policies for all cost ratios. The PdM policy 1, and its optimized version, are characterized by simplicity, fast evaluation, and universal applicability - as long as the prognostics provide the $\text{Pr}(RUL_{\textrm{pred},k}\leq\Delta T)$. Thus, when enough training data are available to optimize the heuristic threshold, this simple PdM policy can outperform other more involved policies. However, in reality, one is typically bounded by the availability of only a limited number of training data, which involves a rather large uncertainty in the estimation of the long-run expected maintenance cost per unit time and the metric $M$. This consequently complicates the task of finding a $p_\textrm{thres}^*$ value that will be optimal also for future components.	

    	\begin{figure}
	\centering
	\begin{subfigure}{.49\textwidth}
		\includegraphics[width=1.\linewidth]{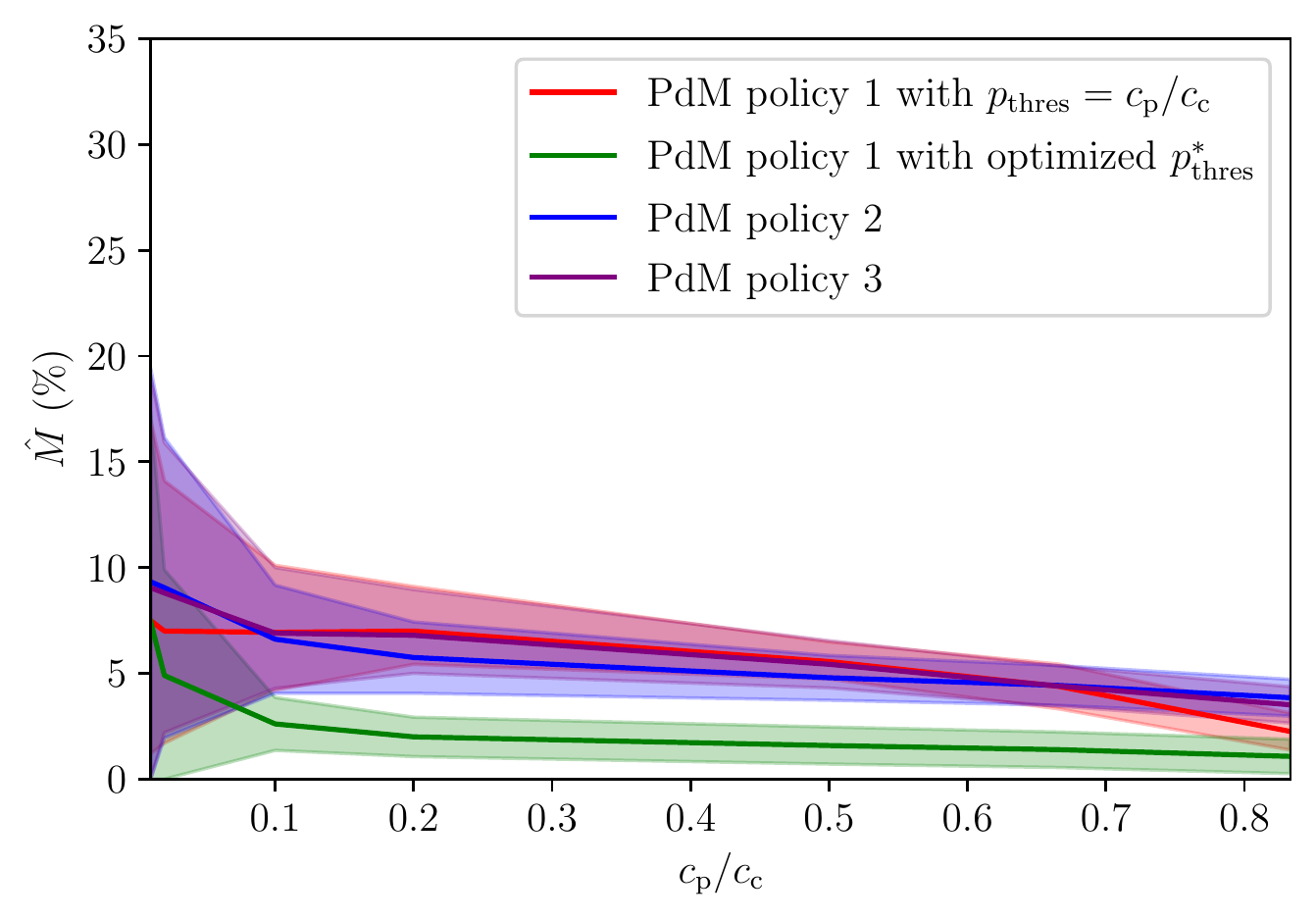}
	\caption{Estimation of the metric $M$ via \cref{eq:metric_hat}, and the associated uncertainty (95\% CIs), in combination with the PdM policies of \cref{subsec:PdM_policy_1}, for different $c_\mathrm{p}/c_\mathrm{c}$ ratios.}
        \label{subfig:4a}
	\end{subfigure}
	\begin{subfigure}{.49\textwidth}
	\includegraphics[width=1.\linewidth]{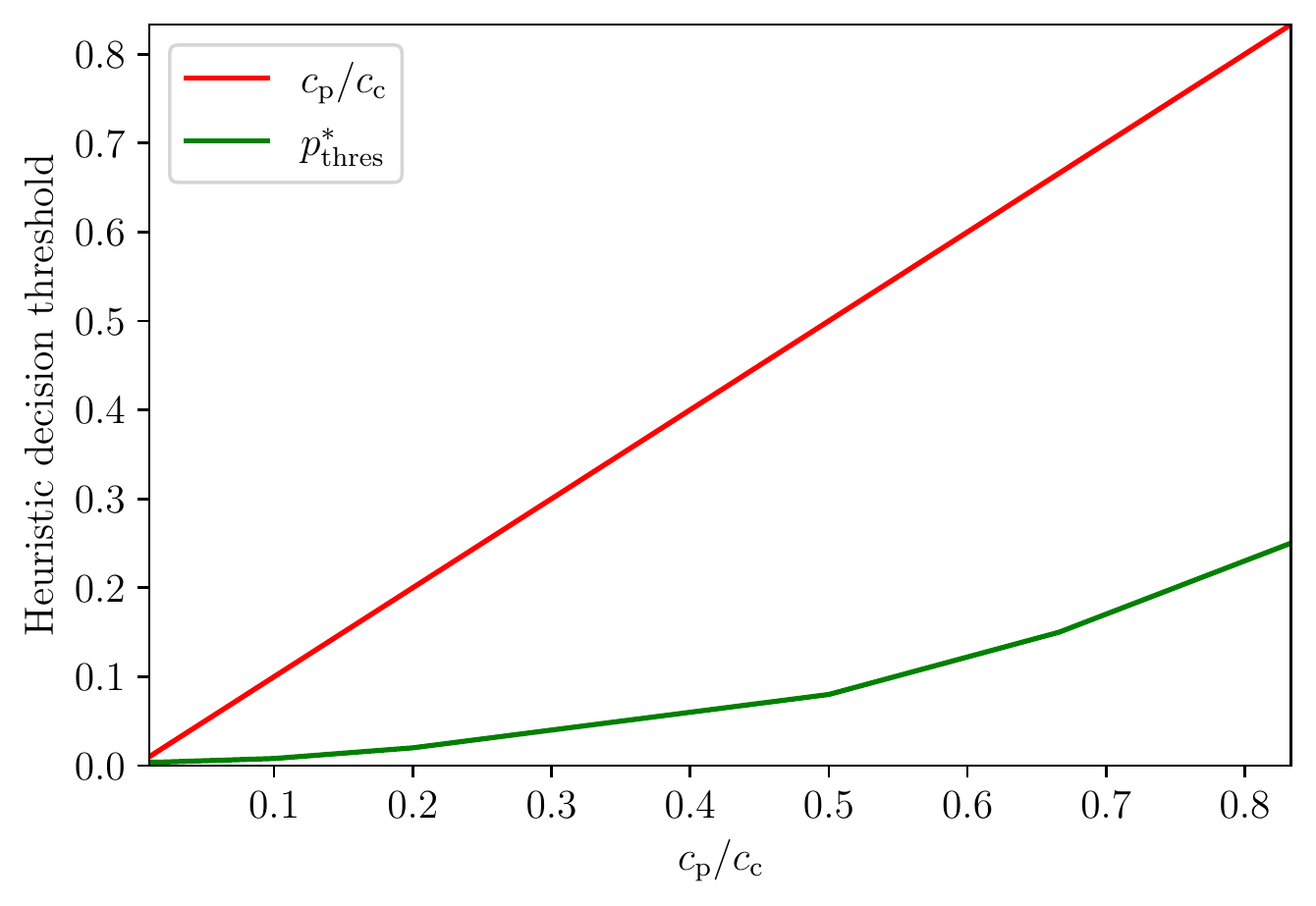}
	\caption{The probability threshold that is used as a heuristic decision threshold in the PdM policy 1. The green line plots the optimal threshold value found for different $c_\mathrm{p}/c_\mathrm{c}$ ratios.}
         \label{subfig:4b}
	\end{subfigure}
        \smallskip
	\caption{Results from virtual RUL simulator with $n=2\cdot10^3$ sampled components for $\sigma_{\text{ln}(\epsilon_k)}=0.15$.}	\label{fig:error_M_generic_1_sigma_015}
	\end{figure}
	
	The uncertainty in the RUL prognostics is propagated to the subsequent PdM planning task. It is expected that reduced uncertainty in the RUL predictions given by a prognostic algorithm leads to enhanced PdM policy performance, and thus to lower values for $M$. This can easily be shown in the context of the virtual RUL simulator. Assuming a smaller value of $\sigma_{\text{ln}(\epsilon_k)}$ corresponds to less uncertain prognostics. \cref{subfig:4a} plots the uncertainty in $\hat{M}$ that we find with each of the employed PdM policies for the same $n=2\cdot 10^3$ sampled components as in \cref{fig:error_M_generic_1}, but for RUL predictions generated with $\sigma_{\text{ln}(\epsilon_k)}=0.15$. It is clear that the values of $\hat{M}$ are significantly reduced compared to the ones reported in \cref{fig:error_M_generic_1}. With less uncertain prognostics, the heuristic PdM policy 1 leads to fewer corrective replacements. At the same time, both PdM policies 2, 3 become less conservative. However, they can lead to corrective replacements even for cases with large $c_\mathrm{c}$ values, as does the heuristic PdM policy 1. This explains the increased uncertainty in quantifying the metric $M$ with these two policies, when compared to \cref{fig:error_M_generic_1}. For small prognostic uncertainty, all three PdM policies lead to comparable results. The optimized heuristic PdM policy 1 again leads to superior performance, further showcasing the benefit of optimizing decision thresholds within simple heuristic PdM policies. \cref{subfig:4b} plots the optimal heuristic threshold $p^*_\textrm{thres}$.

\section{Case study: predictive maintenance of degrading turbofan engines}
\label{sec:Case_study}

In this section we investigate the proposed metric in conjunction with the different PdM policies for comparing different RUL prediction algorithms on a PHM benchmark problem, which involves the degradation simulation of a turbofan engine \cite{CMAPSS}. The data set is publicly available through the NASA Ames Prognostics Data Repository \cite{NASA}.

\subsection{CMAPSS dataset}
Performance degradation histories of a turbofan engine due to wear and tear were numerically generated using the simulator C-MAPSS \cite{Frederick2007}. The engine is simulated under different flight conditions and the effect of performance degradation is introduced in one of the engine modules in the form of an exponential degradation model. The data consists of 14 input variables, which specify the configuration parameters of the simulation, 21 output variables, which provide a measurement snapshot of the response of the system during or after each flight, and 3 variables that describe the operation modes of the engine. The simulations are performed for a number of engine units and they are randomized in the sense that different initial conditions, degradation and noise parameters are selected for each scenario. The dataset contains four data subsets which have been generated with different simulation settings.

At the start of each unit simulation, the engine is normally operating and a fault is introduced in a certain time instance, with the parameters controlling the direction of the failure evolution trajectory randomly selected. The reader is referred to the paper describing the benchmark problem for further details \cite{CMAPSS}. The simulations and the corresponding datasets are split into two parts: i) a training set that contains simulation data, wherein the fault grows in magnitude up to system failure, and ii) a test set that contains simulation data up to a point before system failure. It should be noted that this benchmark dataset has been developed for prediction purposes, with the aim of challenging the development of RUL predictors that are to be evaluated on the test set. However, this study is concerned with the decisions towards optimal PdM planning and as such, the results reported in this section are derived using the FD001 training set, which contains simulation data up to system failure and therefore offers the possibility of exploring the entire decision space. The original training set is split into a training and a test subset, with the former used for the training of prognostic models, as described in the following section, and the latter used for the evaluation of these models with the PdM policies.

For the purpose of the investigations in this paper, the task of PdM planning for each engine unit is simplistically considered as a problem for planning order and replacement actions for a single component. Turbofan engines are complex machinery systems, whose maintenance is in practice planned by taking multiple system-level considerations and logistic constraints into account (e.g., shop loading, parts lead time, consolidated repair planning, etc.). These are not reflected in the simple PdM decision settings considered in this section, but may form part of future investigations.

\subsection{Prognostic models}
\label{subsec: modeling}

Data-driven prognostic models can be considered as mapping functions from a set of input parameters $\mathbf{x}\in\mathbb{R}^{n_{\mathrm{x}}}$, which represent the available system response information, to a set of output quantities of interest (QoI) $\mathbf{y}\in\mathbb{R}^{n_{\mathrm{y}}}$ \footnote{The output QoI $\mathbf{y}$ is typically either some health indicator, which is used for inference of the RUL, or the RUL itself.}, so that $\mathcal{F}: \mathbf{x}\to \mathbf{y}$. This mapping can be expressed by the following mathematical model
\begin{equation}
    \mathbf{y} = \mathcal{F}_{\mathbf{H}}\left(\mathbf{x}, \mathbf{p}\right),
    \label{eq:prognostic-model}
\end{equation}
where $\mathbf{p}$ denotes a set of model-specific parameters, which can be estimated from the training data\footnote{In this work, we choose the mean squared error as the loss function for training regression models, and the log loss function for classification models \cite{Murphy_2012}.} and are used to configure the structure of the underlying model, and $\mathbf{H}$ is the vector of model hyperparameters that are external to the model, and whose values control the training process. The PdM policies presented in \cref{sec:PdM} in this paper rely on the availability of a probabilistic output $\mathbf{y}$ from prognostics. 

The selection of hyperparameters $\mathbf{H}$ for each type of model is typically carried out through an optimization step that minimizes the prediction error evaluated on a set of test or validation data points that are not seen during training. In this work, we instead propose a decision-oriented optimization of hyperparameters:
\begin{equation}
	\mathbf{H}^{*} = \arg\underset{\mathbf{H}}{\min}\,M\left(\mathbf{Y}_k\right),
	\label{eq:hyperparameter-optimization}
\end{equation}
which aims at extracting the optimal hyperparameter configuration that minimizes the proposed metric $M$, which is defined in \cref{eq:metric}, and depends on the model output sequence $\mathbf{Y}_k = \left[ y_1 \ \, y_2 \ \, \ldots y_k\right]$ as well as the corresponding decision setting and adopted PdM policy.

\begin{figure}
    \centering
    \includegraphics[width=0.85\textwidth]{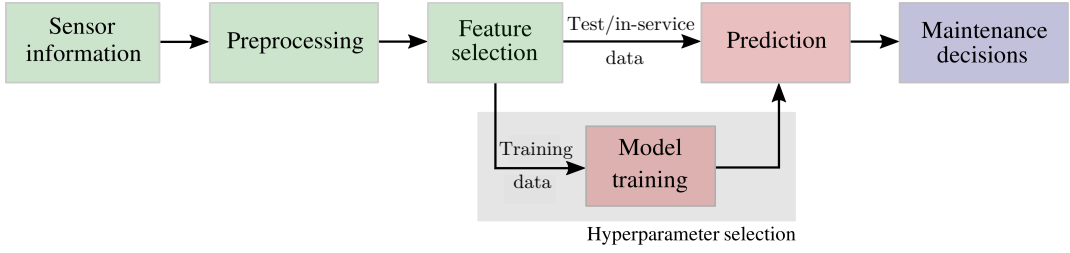}
    \caption{Flowchart of the adopted data-driven predictive maintenance decision process}
    \label{fig:process}
\end{figure}

For the current case study, four different data-driven models are implemented for delivering RUL predictions. Firstly, a Long Short-Term Memory (LSTM) network classifier is constructed in the Keras Python library with the adopted architecture identical to the one proposed in \cite{Nguyen_2019}. The network outputs the class label of the estimated RUL with an associated probability. The second model is a Gaussian naive Bayesian (GNB) classifier, whereby the RUL prediction is again treated as a classification problem \cite{Friedman1997}. In this case, a parent node is used to model the RUL, which can have a healthy or close-to-failure label that is determined on the basis of the conditional probabilities of each sensor signal. The third adopted model is a Decision Trees (DTs) classifier \cite{Breiman1984}, which is estimated using a maximum depth of four so as to prevent overfitting. Finally, the fourth model implements Bayesian filtering of an exponential degradation (EXP) model for performing regression tasks. The EXP regression model relies on the fitting of an exponential model to the first principal component of the sensor data \cite{LESON2013165}. The EXP model parameters are initially extracted by fitting the model to each of the training run-to-failure data sequences, and the sample-based statistical properties of the parameters are used as priors in the prediction phase. Thereafter, the model parameters are sequentially updated upon availability of new sensor measurements, using a Bayesian filtering algorithm \cite{Sarkka_2013}, which delivers the particle-based RUL distribution at each step.

It should be mentioned that the scope of this case study is not the comparison of predictive capabilities of the specific modeling approaches, but the investigation of the metric $M$ and the associated decision policies as a means to compare/evaluate different prognostic algorithms. For the sake of brevity, the reader is referred to the corresponding sources \cite{Nguyen_2019, Zhang2018,Friedman1997,Breiman1984, Saha_2009} for further information on the mathematical background and the theoretical assumptions of each model. 

The adopted steps for PdM planning using all four models are highlighted in \cref{fig:process}. Concretely, the available data from all sensing channels are initially passed through a preprocessing layer, which essentially consists of i) a normalization step, so that all variables are scaled to a standardized range, ii) the labeling of RUL values for the case of classification models, and iii) a smoothing step, which is equivalent to filtering. Thereafter, the features of data to be used as inputs for each model are selected and the training phase is carried out using the training dataset. The remaining data are then used to evaluate the algorithms through the proposed metric $M$. As such, the data-driven prognostic models are trained using an 80\% partition of the FD001 dataset, which contains run-to-failure monitoring data from 100 units, i.e., from 100 different degrading engine modules. For the purpose of our investigations, as explained above, this training set is split into data from 80 units that we use for the training, and data from the remaining 20 units that we employ for the evaluation of the different PdM policies and the metric M.

\subsection{First decision setting: PdM planning for replacement}
 \label{subsec:CMAPSS_first_setting}

For the first decision setting, it is assumed that preventive replacement actions can only be performed at discrete points in time $t_k=k\cdot \Delta T$, for $\Delta T = 10$ flight cycles. The heuristic PdM policy 1 requires the probability $\text{Pr}(RUL_{\mathrm{pred},k}\leq\Delta T)$ as input from the prognostics (see \cref{eq:heuristic1}). During the training process of the LSTM, GNB and DT prognostic classifiers, the output RUL data are labeled into two distinct classes, one corresponding to $RUL>\Delta T$ and the other to $RUL\leq \Delta T$. In this way, the trained classifiers directly output $\text{Pr}(RUL_{\mathrm{pred},k}\leq\Delta T)$ as the associated class probability. With the EXP model, at each time step $t_k$, the $RUL_{\mathrm{pred},k}$ is directly given as output in the form of a vector $\mathbf{y}$ of $n_\mathrm{p}$ weighted samples with a corresponding vector of weights $\mathbf{w}$, obtained via particle filtering \cite{Tatsis_2022,kamariotis_sardi_papaioannou_chatzi_straub_2023}. The required probability is then queried as
\begin{equation}
    \text{Pr}(RUL_{\mathrm{pred},k}\leq\Delta T)=\sum_{i=1}^{n_\mathrm{p}} \mathcal{L}_{0/1}\left(y^{(i)}\leq\Delta T\right)\cdot w^{(i)},
\end{equation}
where $\mathcal{L}_{0/1}$ denotes the 0-1 loss function \cite{Murphy_2012}.

On the other hand, the PdM policies 2 and 3 require prognostic input in the form of the full PDF $f_{RUL_{\mathrm{pred},k}}(t)$. To this end, for all considered prognostic models, some additional post-processing is required. In the case of the EXP model, one simply has to fit the parameters of an appropriately chosen distribution type to the weighted samples $\mathbf{y}$. In the case of the LSTM, GNB, DT prognostic classifiers, given the training routine that we describe in the previous paragraph, the sole prognostic output is the $\text{Pr}(RUL_{\mathrm{pred},k}\leq\Delta T)$, i.e., a single evaluation of the cumulative distribution function (CDF) of $RUL_{\mathrm{pred},k}$. Fitting the CDF of a chosen distribution type to a single available CDF value is challenging. To tackle this problem, for each classifier type, we choose to simultaneously train two classifiers. Let us, for illustration purposes, consider the LSTM model. We train one LSTM model which outputs the $\text{Pr}(RUL_{\mathrm{pred},k}\leq\Delta T)$, and a second LSTM model which outputs another probability, e.g., the $\text{Pr}(RUL_{\mathrm{pred},k}\leq2\cdot\Delta T)$. In such a manner, two values of the CDF are obtained, which enables fitting a chosen distribution type with two parameters. In all models, the lognormal distribution is chosen to model $f_{RUL_{\mathrm{pred},k}}(t)$.

\begin{figure}
\centering
\begin{subfigure}{.49\textwidth}
	\includegraphics[width=1.\linewidth]{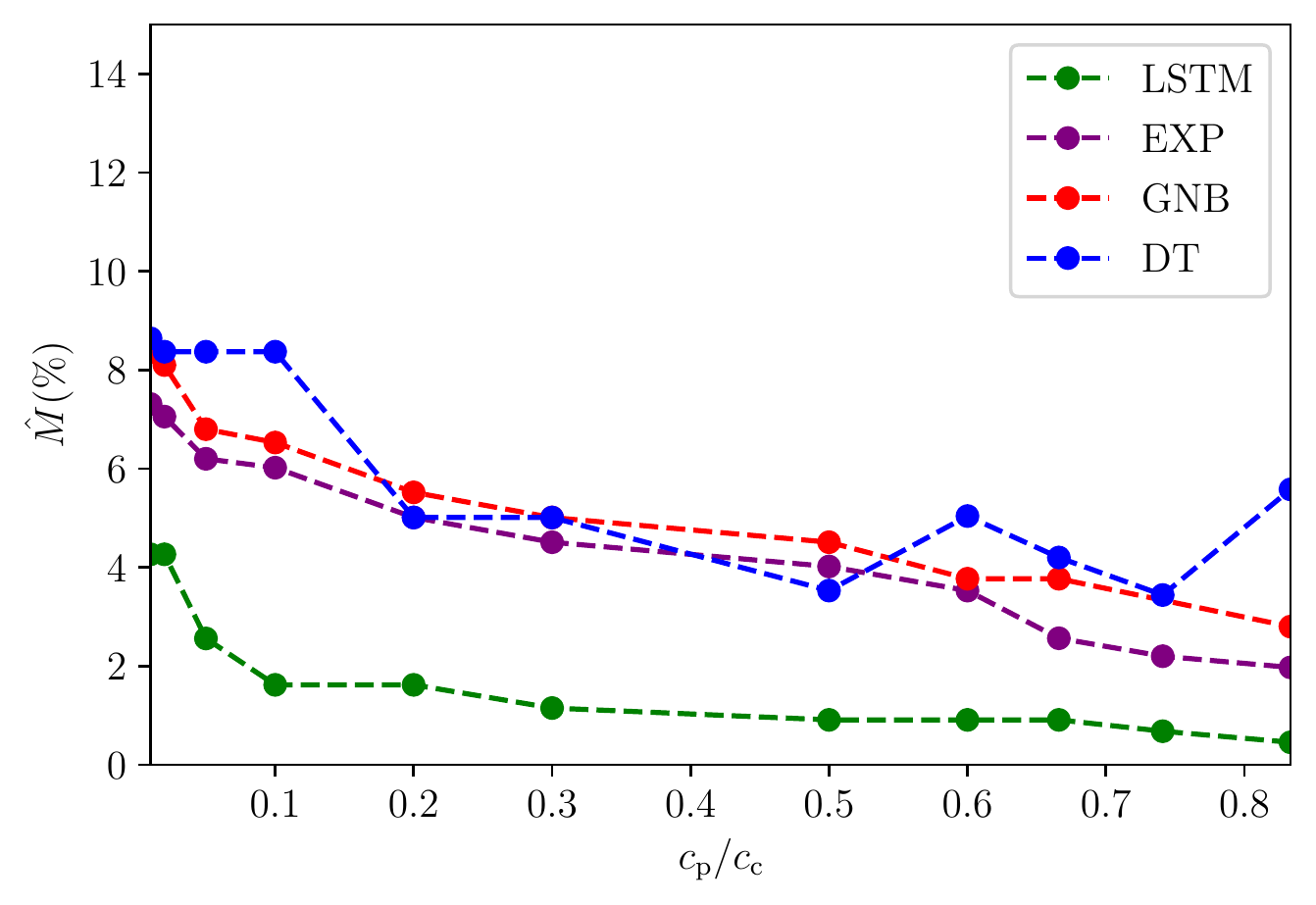}
\subcaption{Heuristic PdM policy 1}
\label{subfig:7a}
\end{subfigure}
\begin{subfigure}{.49\textwidth}
	\includegraphics[width=1.\linewidth]{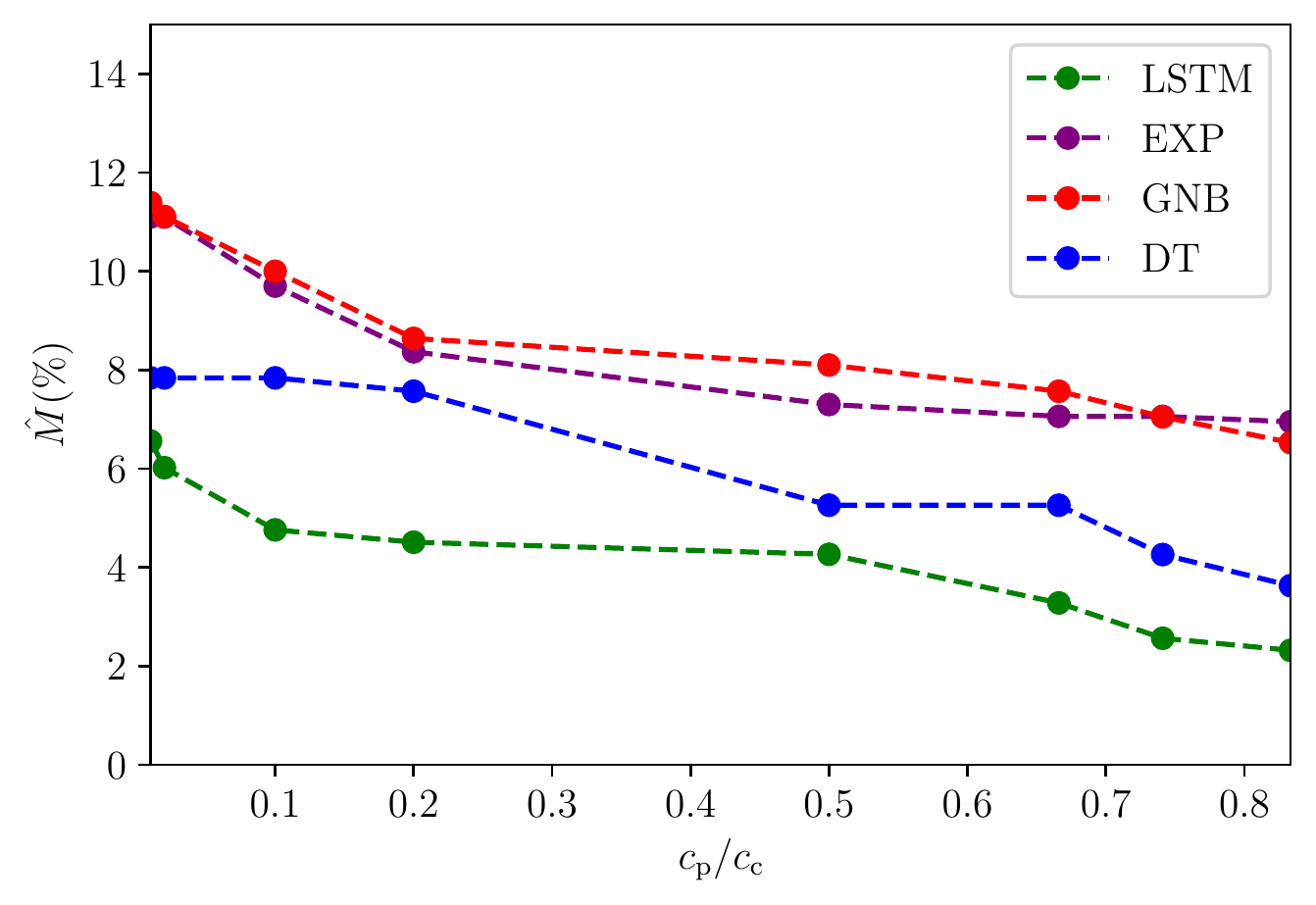}
\subcaption{PdM policy 3}
\label{subfig:7b}
\end{subfigure}
\smallskip
\caption{Evaluation of the metric $M$ in conjunction with a prognostic model and a PdM policy for planning replacement, as a function of different $c_\mathrm{p}/c_\mathrm{c}$ cost ratios. The LSTM prognostic classifier gives the best performance with respect to $M$.}
\label{fig:error_M_CMAPSS}
\end{figure}

\cref{fig:error_M_CMAPSS} plots the results for the metric $M$ computed via the four employed prognostic models for different assumed $c_\mathrm{p}/c_\mathrm{c}$ cost ratios. Specifically, \cref{subfig:7a} plots the results obtained with each model when employing the heuristic PdM policy 1, whereas \cref{subfig:7b} corresponds to the results obtained when employing the here proposed PdM policy 3, which we found to perform better on this dataset compared to the PdM policy 2, which also operates on the basis of the full RUL distribution as input. We thus choose not to include the results obtained with PdM policy 2 in the plot. An initial observation is that, for this specific case study, and for the specific prognostic models, the PdM policy 1 leads to better decisions than the PdM policy 3 for all prognostic models, with the exception of the DT classifier, for which the results are comparable. This result might appear inconsistent with the results obtained in the theoretical investigations of \cref{sec:Generic_example}.

Due to the large uncertainty involved in estimation of $M$ with limited data, a general statement should be made with care. In the current case study, we are limited to 80 available units for training and 20 units for evaluation, which implies presence of significant variability in the results. In particular, evaluation of $M$ on 20 units via \cref{eq:metric_hat} is subject to significant statistical uncertainty. Even though the results in \cref{subfig:7b} appear rather worse than the results in \cref{subfig:7a}, this difference occurs even if the decisions triggered by the two distinct policies are in effect not so different. As an example, let us consider the LSTM model, and the cost ratio $c_\mathrm{p}/c_\mathrm{c}=0.1$. With PdM policy 1 we find $\hat{M}=1.62\%$. This corresponds to preventive replacements informed at $\mathbf{T}_\mathrm{R}$=[230, 200, 290, 260, 180, 260, 170, 200, 210, 150, 130, 330, 150, 250, 280, 330, 190, 150, 180, 190] cycles for the 20 evaluation units and no corrective replacement. Correspondingly, with PdM policy 3 we obtain $\hat{M}=4.76\%$, which corresponds to preventive replacements informed at $\mathbf{T}_\mathrm{R}$=[220, 200, 280, 250, 170,
250, 160, 200, 200, 140, 130, 330, 150, 240, 270, 320, 190, 150, 170, 180] cycles. Comparing the two vectors shows that the difference in $\hat{M}$ originates from the fact that PdM policy 3 informs preventive replacement one decision time step earlier than the PdM policy 1 for some components.

The results in \cref{subfig:7a} reveal that the LSTM prognostic classifier delivers the best performance among all four prognostic models with respect to PdM planning for replacement. The other three models seem to deliver comparable performance. A practitioner could interpret the difference in the results obtained via the different prognostic models in \cref{fig:error_M_CMAPSS} as the percentage of cost savings that using algorithm x for PdM planning could provide compared to using algorithm y. This straightforward interpretation is a good feature for the use of the metric in practice. Naturally, the metric $M$ entirely depends on the choices related to the decision problem, such as, e.g., the values assigned to the costs $c_\mathrm{p}$, $c_\mathrm{c}$. 

Using $M$ as a performance metric has various advantages. Typically, most widely used performance metrics, such as the ones in \cite{Saxena_2010, Nectoux_2012}, or standard metrics such as the MSE of predictions, rely upon a regression type of prognostic outcome. They can therefore not easily compare, e.g., the performance of prognostic regression models directly against prognostic classifiers. This is not the case when using metric $M$, with which any two models can be compared, as long as their prognostic output can be provided as input in the fixed PdM policy. Furthermore, ideally the performance of a prognostic algorithm should be appraised at later prediction stages, when the decisions for preventive replacements actually become relevant, which is what metric $M$ does. Testing a prognostic algorithm with respect to how well it can predict the exact RUL value at an early point in time might provide impractical conclusions. 

\begin{figure}
\centering
\begin{subfigure}{.49\textwidth}
	\includegraphics[width=1.\linewidth]{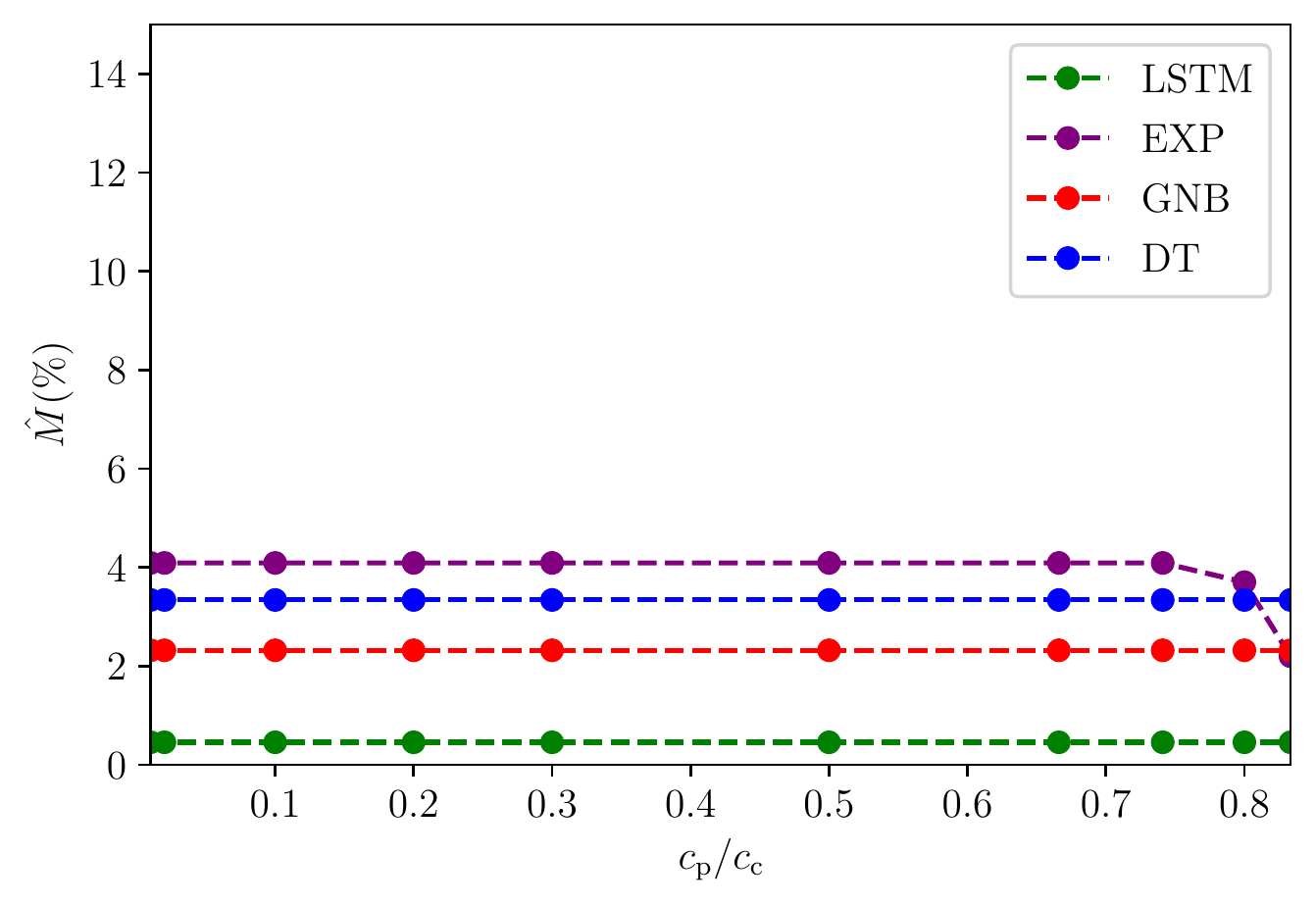}
\subcaption{Value for metric $M$ when employing the PdM policy 1  with optimized $p^*_\mathrm{thres}$.}
\label{subfig:8a}
\end{subfigure}
\begin{subfigure}{.49\textwidth}
	\includegraphics[width=1.\linewidth]{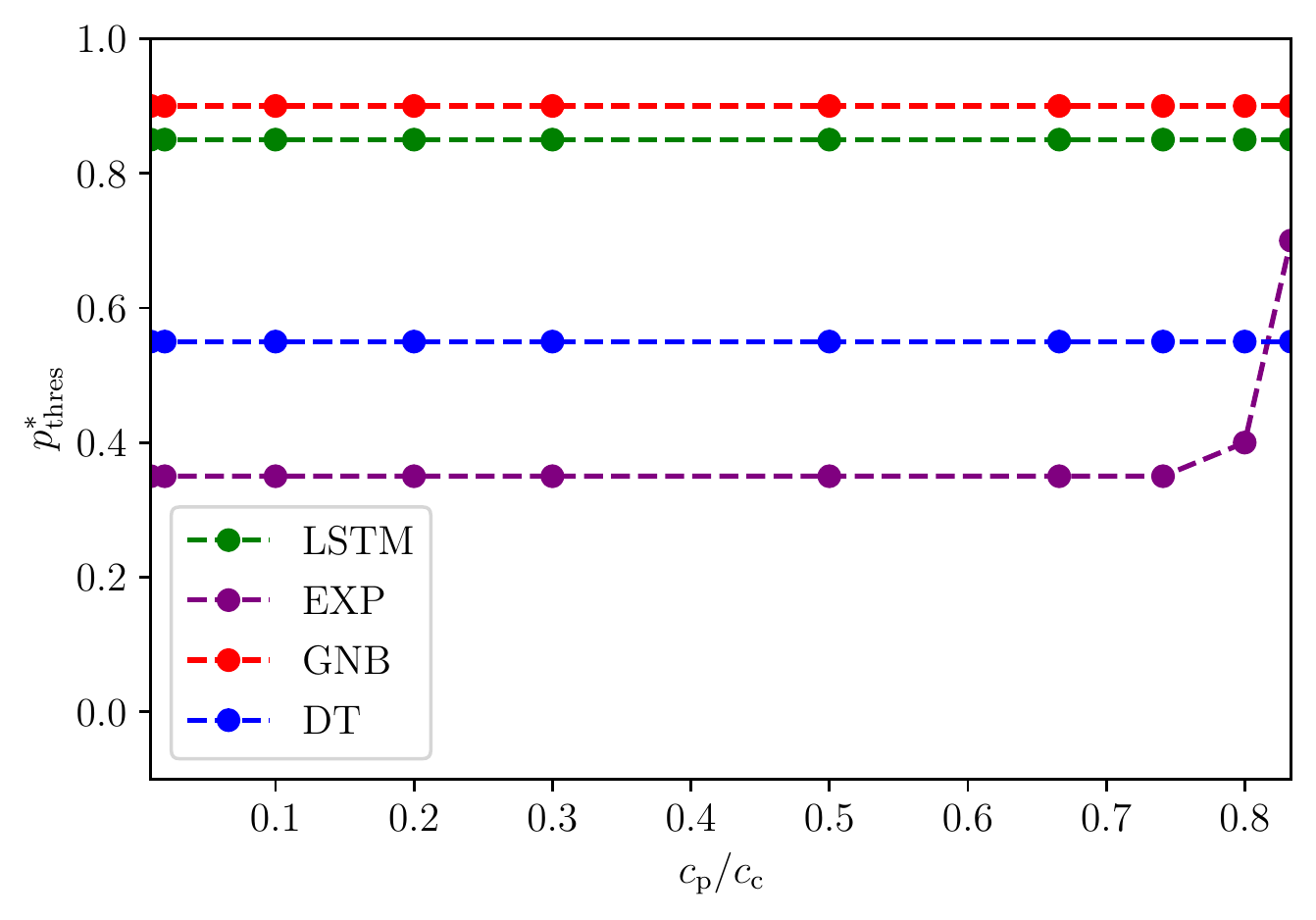}
\subcaption{Optimal heuristic decision thresholds along varying $c_\mathrm{p}/c_\mathrm{c}$ ratios.}
\label{subfig:8b}
\end{subfigure}
\smallskip
\caption{Optimizing the heuristic decision threshold for the heuristic PdM policy 1 within the training process.}
\label{fig:error_M_CMAPSS_opt_heur}
\end{figure}

In \cref{subsubsec:heur_policy_1} we discussed optimizing the heuristic threshold in PdM policy 1, and in \cref{subsec:generic_1} we showed that this can lead to a significant improvement in the PdM decision-making, quantified with respect to metric $M$. This process is also performed in the context of the current case study within the training phase. Specifically, we employ the heuristic decision rule of \cref{eq:heuristic1} and we search for the optimal value $p_\textrm{thres}^*$ that leads to minimization of $\hat{M}$ when evaluating the PdM policy 1 on the 80 training set units. We then employ the heuristic decision rule of \cref{eq:heuristic1} with the optimal value $p_\textrm{thres}^*$ for evaluating $\hat{M}$ on the remaining 20 test set units. The values that the metric $\hat{M}$ assumes for each prognostic model with the corresponding optimized heuristic PdM policy 1 are plotted in \cref{subfig:8a}, demonstrating the non-negligible improvement in the decision-making.

The optimal values of $p_\textrm{thres}^*$ found for different $c_\mathrm{p}/c_\mathrm{c}$ ratios are plotted in \cref{subfig:8b}. It may appear surprising that $p_\textrm{thres}^*$ assumes a large and constant value along the whole $c_\mathrm{p}/c_\mathrm{c}$ axis for most of the models. The reason is that the specific models considered here seem to deliver an overestimation of the classification or regression probabilities. $p_\textrm{thres}^*$ assumes a large value in order to correct for this bias. Furthermore, the fact that the optimization of $p_\textrm{thres}$ is performed on a limited number of units is the reason for which $p_\textrm{thres}^*$ assumes constant values over the $c_\mathrm{p}/c_\mathrm{c}$ axis for most models.

\begin{figure}
\centering
\begin{subfigure}{.49\textwidth}
	\includegraphics[width=1.\linewidth]{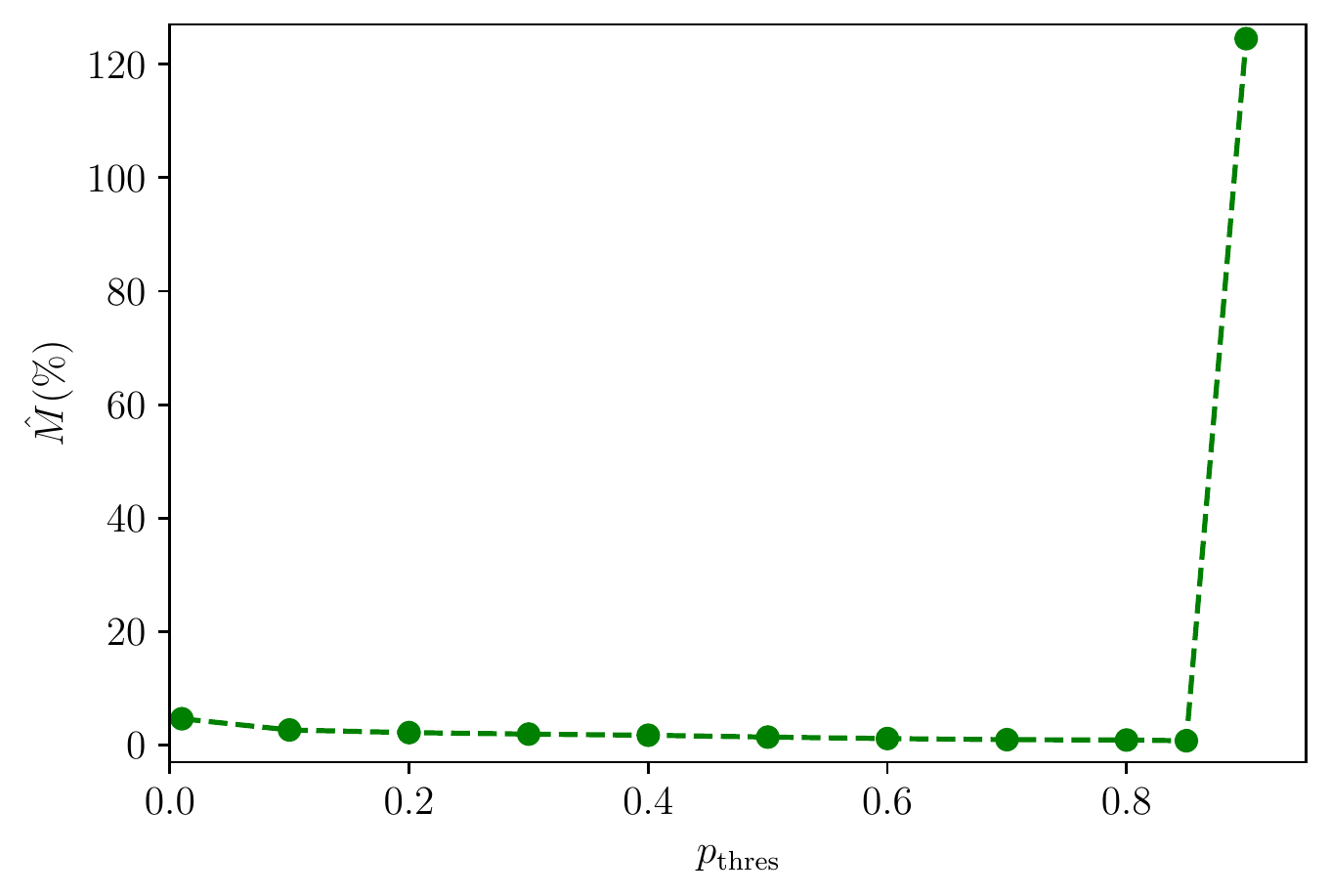}
\subcaption{LSTM model with $c_\mathrm{p}/c_\mathrm{c}=0.01$. For large $c_\mathrm{c}$ values, the risk associated with optimizing $p_\mathrm{thres}$ is large.}
\label{subfig:9a}
\end{subfigure}
\begin{subfigure}{.49\textwidth}
	\includegraphics[width=1.\linewidth]{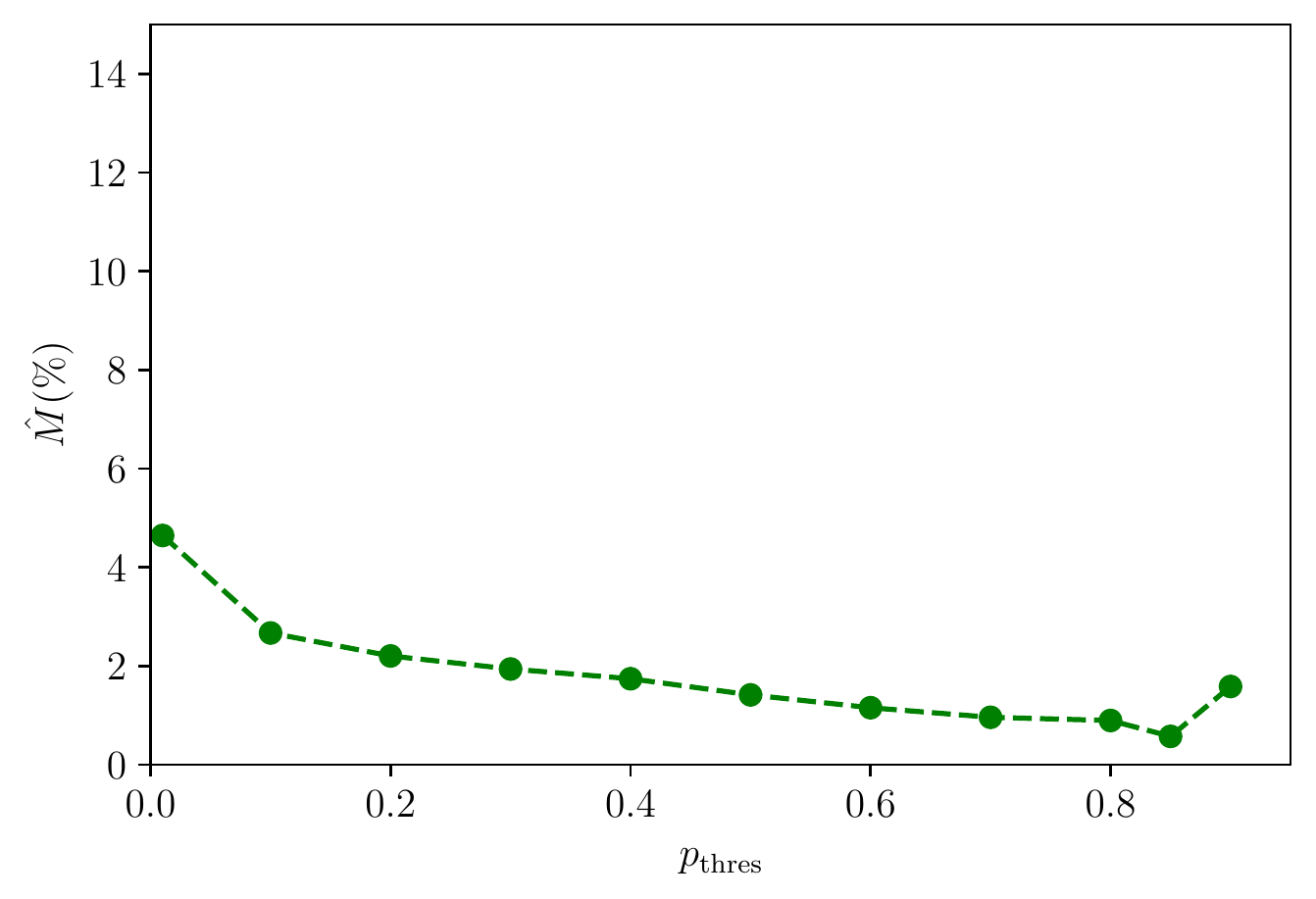}
\subcaption{LSTM model with $c_\mathrm{p}/c_\mathrm{c}=0.5$. For small $c_\mathrm{c}$ values, the risk associated with optimizing $p_\mathrm{thres}$ is small.}
\label{subfig:9b}
\end{subfigure}
\smallskip
\caption{Risk associated with optimizing $p_\mathrm{thres}$ in the heuristic PdM policy 1}
\label{fig:optimize_Pr_thres}
\end{figure}

It should be noted that despite the improvement in the decision-making performance upon optimizing $p_\mathrm{thres}$, a risk always exists in terms of introducing an over-relaxation in the probability space, which can lead to corrective replacements. This might be acceptable for large $c_\mathrm{p}/c_\mathrm{c}$ cost ratio values (e.g., see \cref{subfig:9b}), however, it can lead to significantly poor performance at small cost ratios, where a single corrective replacement is strongly weighted. An illustrative example of such a case is shown in \cref{subfig:9a}. For this studied case, the optimal threshold $p_\textrm{thres}^*=0.85$ is found, which is the value that best accounts for the probability bias that is present in the employed LSTM model. This value results in no triggering of corrective replacements when employing this PdM policy on the 80 training units, and is also the value for which early preventive replacements are minimized. However, a small increase in $p_\textrm{thres}$ (changing its value to 0.9) leads to a very large increase in the value of $M$. This occurs as 2/80 components fail, inducing the very large $c_\mathrm{c}$ cost.

\subsection{Second decision setting: PdM planning for component ordering and replacement}
\label{subsec:CMAPSS_second_setting}
In this section, along with the considerations of \cref{subsec:CMAPSS_first_setting}, we further consider that a component is readily available for replacement only if it was ordered on time. A component should be ordered at a time informed by the heuristic PdM policy of \cref{subsubsec:heur_policy_2}. A deterministic lead time $L=2\cdot \Delta T=20$ cycles is assumed, which is the time from component ordering to delivery. The heuristic PdM policy requires the probabilities $\mathrm{Pr}(RUL_{\mathrm{pred},k}\leq w+\Delta T)$ (see \cref{eq:order_heuristic}) and $\mathrm{Pr}(RUL_{\mathrm{pred},k}\leq \Delta T)$ (see \cref{eq:replacement_heuristic_2}) as input from the prognostics. To this end, during the training process, the output RUL data are labelled into three distinct classes, and the considered prognostic classifiers are trained as multiclass classifiers. 

\begin{figure}
\centering
\begin{subfigure}{.49\textwidth}
	\includegraphics[width=1.\linewidth]{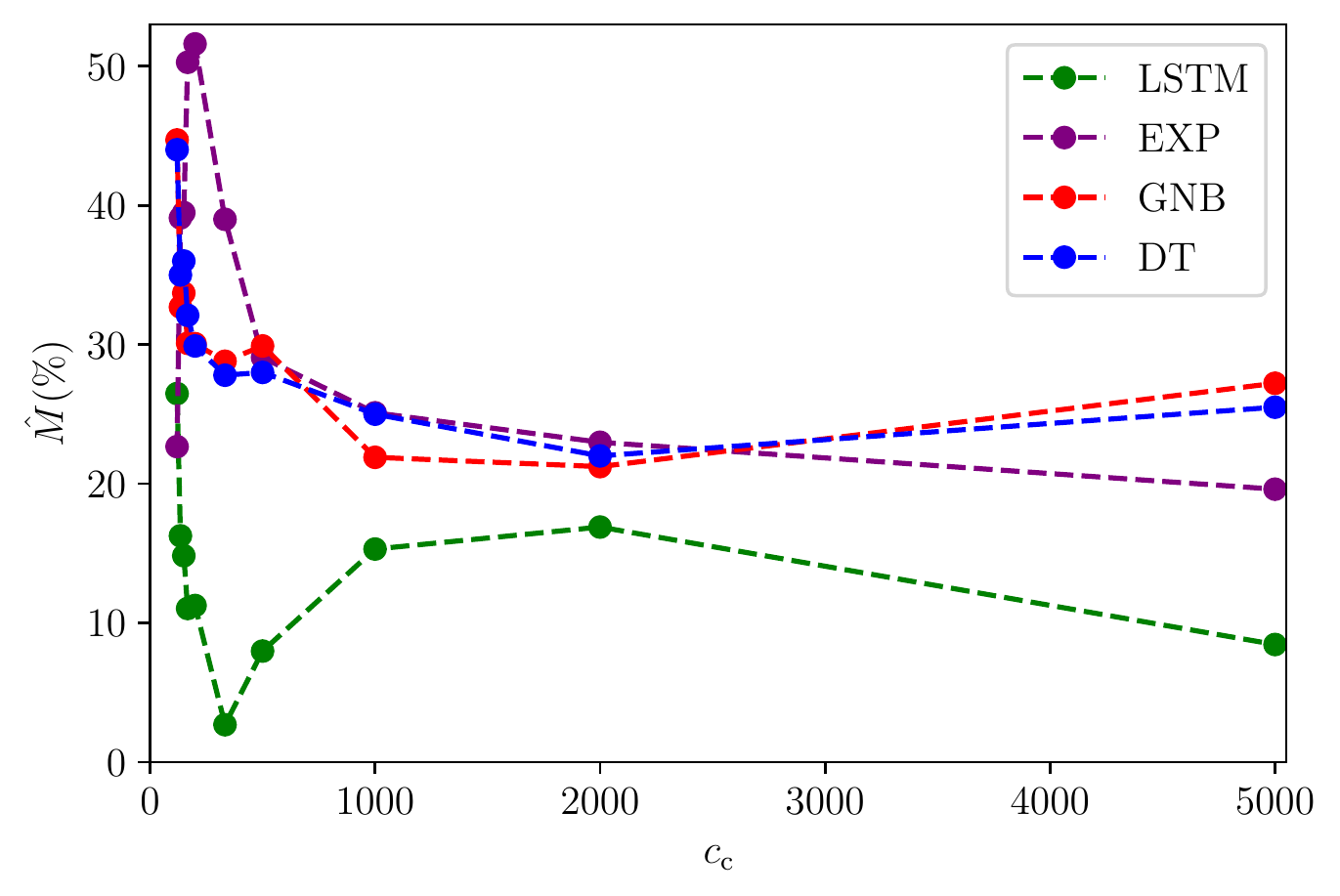}
\subcaption{Choice of threshold values $p^\mathrm{rep}_\mathrm{thres}= p^\mathrm{order}_\mathrm{thres}=c_\mathrm{p}/c_\mathrm{c}$ for the heuristic PdM policy. The LSTM prognostic classifier gives the best performance with respect to $M$.}
\label{subfig:results_second_case_a}
\end{subfigure}
\begin{subfigure}{.49\textwidth}
	\includegraphics[width=1.\linewidth]{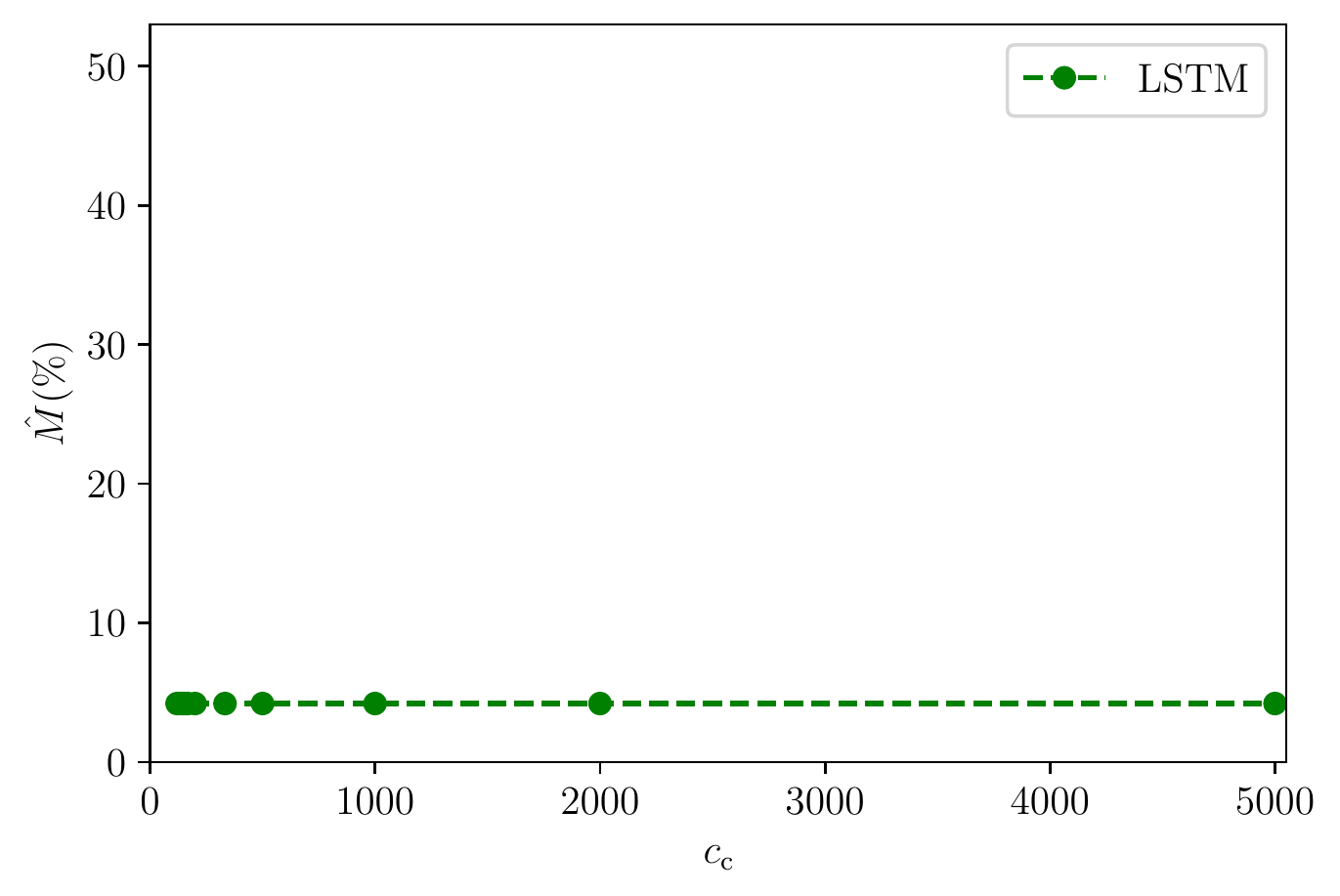}
\subcaption{Heuristic PdM policy with optimal $p^{\mathrm{rep}^*}_\mathrm{thres}=0.5$ and $p^{\mathrm{order}^*}_\mathrm{thres}=0.11$ for the LSTM prognostic classifier.\\}
\label{subfig:results_second_case_b}
\end{subfigure}
\smallskip
\caption{Evaluation of the metric $M$ in conjunction with a prognostic model and the heuristic PdM policy of \cref{subsubsec:heur_policy_2} for planning component ordering and replacement. $\hat{M}$ is plotted as a function of $c_\mathrm{c}$ values varying in the range $[120, 5000]$. The remaining costs are fixed: $c_\mathrm{p}=100$, $c_\mathrm{unav}=10$, $c_\mathrm{inv}=1$.}
\label{fig:2nd_decision}
\end{figure}

\cref{subfig:results_second_case_a} plots the results for the metric $M$ computed via the four employed prognostic models for varying $c_\mathrm{c}$ cost, fixed costs $c_\mathrm{p}=100$, $c_\mathrm{unav}=10$, $c_\mathrm{inv}=1$ and heuristic threshold values $p^\mathrm{rep}_\mathrm{thres}= p^\mathrm{order}_\mathrm{thres}=c_\mathrm{p}/c_\mathrm{c}$. The LSTM prognostic classifier delivers the best performance with respect to PdM planning for component ordering and replacement, with the other three models delivering comparable performance. For all four models, the metric $M$ assumes significantly higher values than those of \cref{subfig:7a}, owing to the additional costs related to late ordering and holding inventory of a component. Naturally, the magnitude of $M$ strongly depends on the chosen $c_\mathrm{p}, c_\mathrm{c}, c_\mathrm{unav}, c_\mathrm{inv}$ costs.

The heuristic decision rule of \cref{eq:replacement_heuristic_2} determines the PR versus DN action without taking into account whether or not a new component is in stock. This proves to be a suboptimal choice, especially when the $c_\mathrm{unav}$ value is non-negligible. Let us take the LSTM model and the engine unit with ID=100 (with true failure time at 200 cycles) as an example, for $c_\mathrm{c}=1000$. The heuristic PdM policy informs component ordering at $T_\mathrm{order}^{(i)}=170$ cycles, and a preventive replacement already at the next decision time step, i.e., at $T_\mathrm{R}^{(i)}=180$ cycles with component unavailability, which based on \cref{eq:cost_delay} induces $C_\mathrm{delay}^{(i)} = 100$. Hence, this simple heuristic PdM policy should be improved in the future.

For the LSTM model, we additionally perform an optimization of the two heuristic thresholds on the training data. We then employ the heuristic PdM policy with the optimal values $p^{\mathrm{order}^*}_\mathrm{thres}=0.11$ and $p^{\mathrm{rep}^*}_\mathrm{thres}=0.5$ for evaluating $\hat{M}$ on the test set units. The results plotted in \cref{subfig:results_second_case_b} demonstrate that this threshold optimization leads to a significant improvement of the policy.

\section{Concluding remarks}
\label{sec:Conclusions}

In this paper, we introduce a decision-oriented metric $M$ for assessing and optimizing data-driven prognostic algorithms. The proposed metric assesses and optimizes algorithms by accounting for their effect on downstream predictive maintenance (PdM) decisions that are to be triggered by their predictions (outputs). Hence, it is defined in association with a specific decision context and a corresponding PdM policy, which informs the maintenance actions based on input uncertain Remaining Useful Life (RUL) predictions. Here, we specifically define and discuss the metric within two common PdM decision settings: i) component replacement planning and ii) component ordering-replacement
planning. We numerically investigate the metric with the aid of: 1) a hypothetical virtual RUL simulator and 2) an application case study related to turbofan engine degradation, for which a run-to-failure dataset is readily available (the CMAPSS dataset). For the latter case study, four data-driven prognostic models for classification and regression are employed. We tune the hyperparameters of these algorithms and assess their performance on the basis of the decision-oriented metric $M$. 

For component replacement planning, we discuss two PdM policies of varying complexity that are most commonly used in the PHM literature. The first policy is a simple heuristic policy, which informs replacement via imposing a heuristic threshold on the probability of RUL exceedance at the next decision time step. A significant improvement to this policy occurs when optimizing the value of the heuristic threshold. The optimal value is found as the argument that minimizes the metric $M$ estimated on $n$ run-to-failure experiments contained in the training dataset. The second policy operates on the basis of the availability of the full RUL distribution, and searches for the optimal future time to replacement. This is done with the aid of a renewal-theory-based objective function. This objective function, which is derived from an assumption adopted in state-of-the-art literature, incorrectly assumes that the predicted distribution of the time to failure of the component corresponds to the underlying distribution of the time to failure of the whole population of components. We here clarify this, and propose an alternative objective function based on an alternative assumption, which is shown to lead to an enhanced performance. 

For component ordering-replacement planning, we restrict ourselves to one simple heuristic PdM policy, which informs ordering and replacement via heuristic thresholds on the probability of RUL exceedance in future decision time steps. We show that optimizing the value of the heuristic thresholds leads to a considerable improvement of this PdM policy.

For the CMAPSS case study, the Long Short-Term Memory (LSTM) network classifier is shown to deliver the best performance among the implemented prognostic models with respect to both PdM for replacement planning and PdM for component ordering-replacement planning.

The proposed decision-oriented performance assessment of prognostic algorithms through the metric $M$ has multiple advantages over conventional metrics that appraise the quality of a prognostic algorithm. Notably, $M$ can be used to compare the performance of any two algorithms, i.e., it does not require a regression-based prognostic outcome like most prediction-based metrics. Besides, $M$ automatically appraises the efficacy of prognostic algorithms at later prediction stages, which are crucial for decision-making. Finally, the metric $M$ can be interpreted as the percentage cost savings for maintenance associated with the use of each algorithm relative to the perfect policy.

The availability of monitoring datasets from run-to-failure experiments is essential to a data-driven evaluation of the proposed metric. This could potentially free the analyst altogether from the need of \textit{a-priori} defining a stochastic model describing the deterioration process. Availability of only a limited amount of such data, however, poses a bottleneck for this evaluation, as it leads to an estimate with fairly large variability. For instance, in the CMAPSS case study, we see that 20 run-to-failure samples are not sufficient for obtaining a reliable estimate of the metric $M$. Furthermore, the metric depends on the initial choice of different cost values associated with a decision setting, e.g., the preventive/corrective replacement cost, the unavailability cost. In order to perform and optimize maintenance planning, one cannot escape quantifying these costs. While an exact estimate of these uncertain costs is often difficult to obtain in practice, a rough estimate can typically be made be experienced engineers. For the purpose of defining the metric, such rough estimates should be sufficient for practical applications (see \cref{fig:error_M_CMAPSS,fig:2nd_decision}), considering that alternative metrics do not consider and account for these costs at all.

A promising avenue of future research relates to training of prognostic algorithms to receive monitoring data as input and directly output a decision within a certain decision setting, whereby the PdM policy can be learnt during the training process, e.g., via deep reinforcement learning \cite{ANDRIOTIS2021107551,LEE2023108908}. Such advanced policies will typically need to be calibrated to the specifics of the cost model, deterioration processes and monitoring data, and would require the availability of a large amount of training data.

\section*{CRediT authorship contribution statement}
\textbf{Antonios Kamariotis}: Conceptualization, Methodology, Software, Formal analysis, Visualization, Writing - original draft.
\textbf{Konstantinos Tatsis}: Methodology, Software, Visualization, Writing - original draft.
\textbf{Eleni Chatzi}: Conceptualization, Methodology, Writing - review \& editing, Funding acquisition.
\textbf{Kai Goebel}: Writing - review \& editing.
\textbf{Daniel Straub}: Conceptualization, Methodology, Writing - review \& editing, Funding acquisition.

\section*{Declaration of competing interest}
The authors declare that they have no known competing financial interests or personal relationships that could have appeared to influence the work reported in this paper.
	
\section*{Acknowledgments}
The work of A. Kamariotis and E. Chatzi has been carried out with the support of the Technical University of Munich - Institute for Advanced Study, Germany, funded by the German Excellence Initiative and the T{\"U}V S{\"U}D Foundation.
	
\bibliography{mybibfile}
	
\end{document}